\documentclass{aa}  

\usepackage{graphicx}
\usepackage{txfonts}
\usepackage{caption}

\begin{document}

   \title{NuSTAR view of Swift/BAT AGN: The $R$-$\Gamma$ correlation}

   \author{C. Panagiotou
          \and
          R. Walter
          }

   \institute{
              Astronomy Department, University of Geneva, Chemin d’Ecogia 16, 1290 Versoix, Switzerland\\
              \email{Christos.Panagiotou@unige.ch}
         }

   \date{Received ***; accepted ***}

 
  \abstract
   {The reflection hump is a prominent feature in the hard X-ray spectrum of active galactic nuclei (AGN). Its exact shape and its correlation to other quantities provide valuable information about the inner and outer regions of an AGN.}
   {Our main goal is to study the reflection hump in a large sample of nearby AGN. We aim to investigate the evolution of reflection with absorption and its correlation with the spectral index.}
   {We analysed archived \textit{NuSTAR} observations of the 70-month BAT catalogue AGN. By performing a detailed spectral analysis, we were able to constrain the spectral parameters and to investigate the reflection emission in a large sample of individual sources.}
   {The reflection strength was found to be strongly correlated with the power-law slope in unabsorbed sources, pointing towards disc reflection for these sources. Different possible explanations were tested and the most likely one is that the corona is moving either towards or away from the disc with a moderately relativistic velocity. An $R-\Gamma$ correlation was not detected for absorbed sources. In addition, these AGN feature harder spectra, suggesting intrinsic differences between the two classes or a slab geometry for the X-ray source.  }
  {}


\maketitle


\section{Introduction}

The X-ray emission of Active Galactic Nuclei (AGN) is thought to originate from a compact region located close to the central supermassive black hole. According to the currently accepted model, UV and optical photons emitted by the accretion disc enter a region filled with high-energy electrons, often referred to as corona, and are upscattered to X-rays by inverse Compton scattering. Part of this primary X-ray emission is then intercepted by the surrounding material and reflected towards our line of sight. 

The spectrum of thermal Comptonisation in an optically thin region is approximately represented by a cut-off power law, which has been found to aptly describe the observed X-ray spectra of AGN \citep[e.g.][]{1993ApJ...413..507H}. In addition, a distinct reflection feature is commonly observed. The so-called Compton hump is an excess of emission in comparison to a simple power-law peaking at $\sim$30 keV. Finally, a fluorescent iron line at $\sim$6.4 keV is prominent in the spectra of most AGN. This line is also produced by the interaction of the primary X-ray emission with the surrounding material.

According to the unification model of AGN \citep{1993ARA&A..31..473A}, a toroidal dusty region surrounds both the corona and disc structure. The so-called torus was initially proposed to account for the lack of broad emission lines in the optical spectrum of Type 2 AGN. If it lies in our line of sight, the gas in the torus will also absorb part of the X-ray emission. In addition to producing absorption effects, the torus is expected to contribute significantly to the reflected emission, together with the disc.

Although the above model is successful in explaining the broad characteristics of AGN X-ray emission, there are still many open questions that require more careful treatment. Several works have followed different techniques with the aim of improving our understanding of the central X-ray source and its surrounding geometry.

Time variability and microlensing studies \citep[e.g.][]{2013ApJ...769...53M} have been successful in constraining the X-ray source position within a few tens of gravitational radii from the black hole. More recently, the development of reverberation techniques allowed to study the regions in the immediate vicinity of the black hole. These studies have put stronger constraints on the position of the corona with respect to the black hole and have also identified the inner disc as the main source of the Fe line and Compton hump emissions for a handful of sources \citep[e.g.][]{2014ApJ...789...56Z, 2015MNRAS.446..737K}.

Another commonly followed approach to study the AGN X-ray emission is the investigation for correlation amongst the various spectral parameters or physical parameters. Such correlations are expected to be driven by the underlying physical processes or the exact geometry of the source and could, therefore, provide additional information. 

\cite{2013MNRAS.433.2485B} have found that the power-law slope is positively correlated to the Eddington ratio, suggesting that the accretion rate determines the physical conditions of the corona. The situation is more complicated when the dependence of the slope on the X-ray luminosity is explored. Some studies \citep[e.g.][]{2008AJ....135.1505S} have suggested the existence of a positive correlation, while others \citep[e.g.][]{2011MNRAS.417..992S} observed the opposite trend. On the other hand, studies based on a local sample have found no evidence of a significant correlation \citep[e.g.][]{2009ApJ...690.1322W}. 

The reflection strength, R, which is a parametrisation of how strong the reflected emission is with respect to the primary power-law, was also found to be correlated to the X-ray slope. \cite{1999MNRAS.303L..11Z} were the first to observe a positive correlation between the two parameters for a group of Seyfert galaxies, which they interpreted as a result of the interplay between the disc and the corona. Although the robustness of this correlation was questioned by \cite{2001ApJ...548..694V}, \cite{2003MNRAS.342..355Z} studied in detail the various systematic and statistical effects and confirmed the reality of the above correlation.  More recently, \cite{2018ApJ...854...33Z} found an anti-correlation between R and the X-ray luminosity.

In \cite{2019A&A...626A..40P}, we studied the X-ray properties of a sample of nearby Seyfert galaxies using the Nuclear Spectroscopic Telescope Array \citep[NuSTAR,][]{2013ApJ...770..103H}, the first telescope focusing X-rays above 10 keV, which makes it ideal to study the reflection hump in individual AGN. We found that the reflection emission varies with absorption, which potentially points to a different origin of reflection between absorbed and unabsorbed sources. Reflection was found to correlate with the X-ray slope in unabsorbed sources, while evidence for a correlation between the observed absorption and the reflection was found for the obscured sources.

Motivated by these results, we expanded our analysis to a much larger sample. Our main goal was to study the hard X-ray spectrum of AGN and investigate for differences between absorbed and unabsorbed sources. In this work, we present the analysis of non Compton thick sources, focusing mainly in the observed $R$-$\Gamma$ correlation and its interpretation. We describe the considered sample and the applied reduction in Sect. \ref{sec:data_red}. The spectral analysis followed in this study is discussed in Sect. \ref{Sec:spec_anal}. The results are presented in Sect. \ref{sec:results} and are discussed in further detail in Sect. \ref{sec:discuss}. Our main findings are summarised in Sect. \ref{sec:conclude}.

\section{Data sample and reduction}
\label{sec:data_red}

Our sample comprises the sources of the 70-month Swift-BAT catalogue \citep{2013ApJS..207...19B} that were observed by NuSTAR and which had public archival data until April 2019. We considered the sources that are categorised as Class 4 and 5 in the BAT catalogue, which correspond to the Type 1 and 2 Seyfert galaxies, respectively. In total, there were 128 Seyfert 1 and 168 Seyfert 2 objects. 

We followed the standardised procedure in order to reduce the observational data. We used the NuSTAR Data Analysis Software (NuSTARDAS) package to produce clean event files. Due to the passage of NuSTAR through the South Antlantic Anomaly (SAA) area, some events were removed when needed. The source spectra were extracted from a circular region centered on the source's celestial coordinates and the background spectra from a source-free annulus surrounding the source region. When the source was located close to the detector's edge or close to another source, the background spectrum was extracted from a nearby source-free circular region. The source region's radius was determined by a visual examination of the source's image with the aim to maximise the signal-to-noise ratio, while the inner radius of the background annulus was always defined to be at least 30 arcseconds larger than the source's radius in order to avoid source contamination. Finally, all the source spectra were binned with at least 25 source counts per bin.

We, then, examined the spectral variability of all sources with more than one observation. When the source spectral shape was found to be not significantly variable, we calculated its average spectrum using the \textit{addspec} tool. A spectral variability is considered as non-significant when an initial fit of the different observations reveals that the spectral parameters are consistent within the errors between the observations, with a potential exception for the normalisation. In other words, we estimated an average spectrum when the spectal shape of the source remains the same, while its flux state might be variable. If the source spectrum was moderately variable, we considered the spectrum of each individual observation separately.
Finally, there were a few sources that underwent severe variability between their observations, showing a transition from a reflection dominated to a continuum dominated spectrum. These sources were omitted from the subsequent analysis.

Several more sources were excluded based on various criteria. First, we excluded the sources with a low signal-to-noise ratio. Only the sources with $S/N > 31$ \footnote{The signal-to-noise ratio is calculated using the spectrum of both detectors FPMA and FPMB.}, which was found to be sufficiently high for the spectral parameters to be constrained, were considered in this study. Second, we excluded the sources featuring a reflection dominated spectrum. Initially, we pursued a model-indepedent approach, calculating the following softness ratio:

\begin{eqnarray}
    SR &= \frac{CR_{3-5} - CR_{25-35}}{CR_{3-5} + CR_{25-35}},
\end{eqnarray} 

\noindent where $CR_{3-5}$ denotes the count rate from 3 to 5 keV and $CR_{25-35}$ the count rate from 25 to 35 keV. We have found that the sources with $SR<0.2$ have a reflection dominated spectrum and are more likely to be Compton thick \citep[][where all sources categorised as Class 5 have $SR<0.2$]{2019A&A...626A..40P}. Therefore, we decided to exclude the sources with $SR<0.2$ from our current analysis. It should be noted that using a different energy range for the hard X-rays (say 15-25 keV) does not modify our results. In addition, we excluded some sources for which the spectral model discussed in Sect. \ref{Sec:spec_anal} required a large amount of reflection (reflection parameter $R>4$) to be well fitted. All these sources, with an apparent reflection dominated spectrum, require a complex spectral model.

We also excluded the galaxies that have been found to host a LINER, as well as \textit{SWIFT J0319.7+4132}, which lies in the center of Perseus cluster. All the sources that have been excluded are listed in Table \ref{tab:sources_exclud}. 

After the exclusion of sources, the considered sample consists of 113 Seyfert 1 and 101 Seyfert 2. Nearly a fourth of them (24 Seyfert 1 and 27 Seyfert 2) have already been studied in \cite{2019A&A...626A..40P}, where we retrieved the results from. For the new sources, we followed the spectral analysis outlined in the next section. This analysis and the corresponding assumptions are the same as these followed in \cite{2019A&A...626A..40P}. The objects' characteristics and their observational details are listed in Tables \ref{tab:sources_info} and \ref{tab:obs_log}, respectively. The sources denoted by a dagger in the former table have been analysed in \cite{2019A&A...626A..40P}. 

\begin{figure}
  \centering
  \includegraphics[width=0.49\linewidth,height=0.56\linewidth, clip]{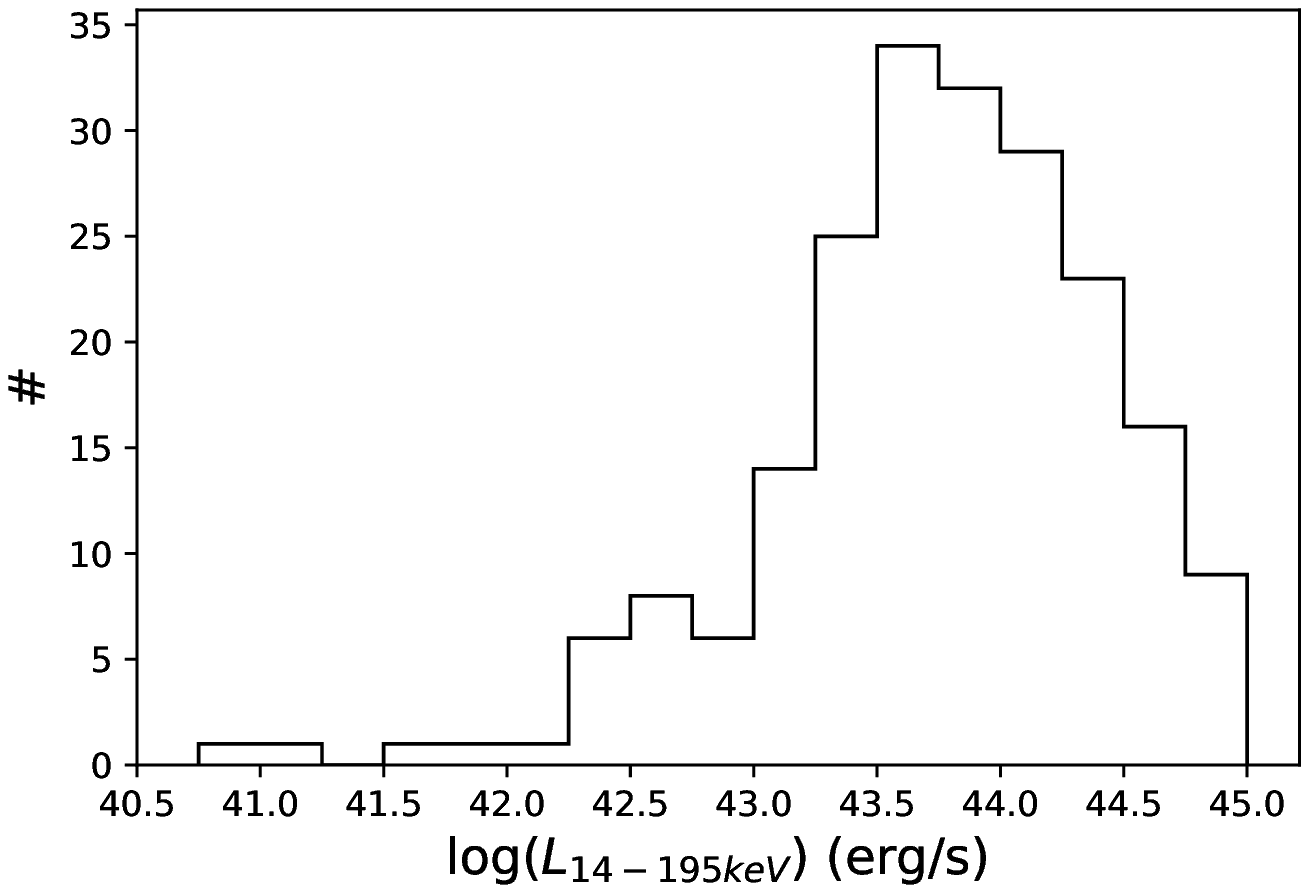}
  \includegraphics[width=0.49\linewidth,height=0.56\linewidth, clip]{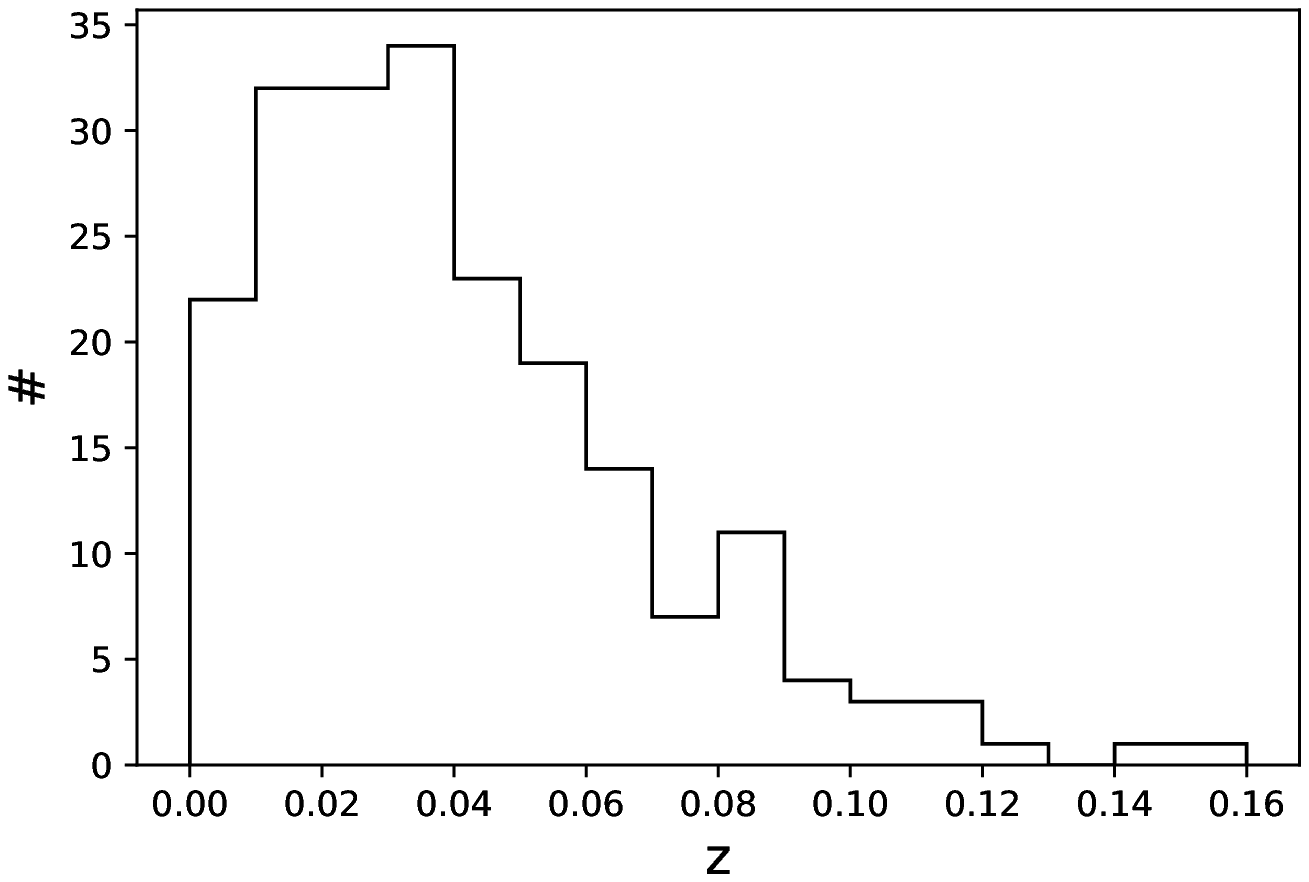}
  \caption[10]{Distribution of the 14-195 keV luminosity (left) and redshift (right panel) of the used sources, as given in the BAT catalogue. }
  \label{fig:bat_lumred}
\end{figure}

The used sample spans a range of luminosities and redshifts. Figure \ref{fig:bat_lumred} plots the distribution of redshift and BAT 14-195 keV observed luminosity.

\section{Spectral Analysis}

\label{Sec:spec_anal}

\subsection{Baseline model}
\label{Sec:base_model}

The spectra of all sources were fitted by the same model. The spectra of the two NuSTAR detectors, FPMA and FPMB, were considered together allowing for a cross-normalisation constant to be minimised. The XSPEC software \citep{1996ASPC..101...17A} was used for the fitting procedure and the element abundances of \cite{2009ARA&A..47..481A} were assumed. We performed the model fitting using $\chi^2$ statistics. All the errors reported in this work correspond to 1-$\sigma$ confidence interval, unless otherwise noted. We have assumed a $\Lambda$CDM cosmology of $H_0 = 70 \text{km}/ \text{s}/ \text{Mpc}$ and $\Omega_\Lambda = 0.73$ .

We used the phenomenological model \textit{pexrav} \citep{1995MNRAS.273..837M} to simulate the AGN emission as a power-law with an exponential cut-off at high energies plus a reflection component produced by the scattering of the power-law emission in a neutral medium with an infinite optical depth. This emission is modified by photoelectric absorption, modelled by \textit{zphabs} in XSPEC, due to the existence of gas in the host galaxy. The absorption due to the Galactic interstellar medium was not considered. Towards most of the sources the Galactic absorption column density is well below $10^{22} \text{cm}^{-2}$, and thus has an insignificant effect on the NuSTAR spectra, which start at 3 keV, even for \textit{SWIFT J1347.4-6033} and \textit{J2018.8+4041}, for which $N_{H,Gal} = 0.96$ and $1.1 \cdot 10^{22} \text{cm}^{-2}$, respectively.

Finally, we used a gaussian emission line centered at 6.4 keV, to account for the Fe K$\alpha$ line. To simplify the fit we assumed the existence of only one narrow Fe line, with a width fixed at 0.05 keV, unless it was otherwise needed by the fit (Sect. \ref{Sec:fe_line}). We, furthermore, assumed the same inclination angle for all the sources, $\text{cos } i = 0.45$, and solar abundances for the reflecting and absorbing media. There were seven model parameters left to be minimised during the fit, that is the intrinsic absorption column density, $N_H$, the power-law index, $\Gamma$, the reflection strength, $R$, the energy of the cut-off, $E_C$, the two normalisations of the iron line and of the power-law and the cross-normalisation between the two detectors. 

The above model was initially used to fit all the observed spectra and provided a statistically accepted fit for the majority of the sources. There were several sources, for which a more careful treatment of the Fe emission was needed. These objects are discussed in detail in Sect. \ref{Sec:fe_line}. In addition, seven sources with multiple observations were found to exhibit a moderate spectral variability between the observation periods. We fitted the individual spectra of each observation for these objects. These sources have been excluded from the subsequent analysis of the different classes (Sect. \ref{Sec:class}).

Furthermore, the best-fit $\Gamma$ for thirteen sources was found to be below $1.4$. Such a small value is unphysical and not commonly observed in the spectra of AGN. These small values might be an artifact of the fit. The spectral shape of a highly absorbed source might be equally well reproduced by a less absorbed low-$\Gamma$ power-law with a  small energy cut-off and by an absorbed power-law of larger $\Gamma$, $E_C$ and $R$. Therefore, we fixed $\Gamma$ to a value of 1.73\footnote{This is the average best-fit $\Gamma$ of the MOB class (Sect. \ref{Sec:class}), to which all these sources belong.} and repeated the fit for these sources. The new fits are still statistically accepted. Although we give the results for these sources, we excluded them from the subsequent discussion to avoid introducing biases in our results.

Table \ref{tab:best_fit} lists the best-fit results for all sources. The last column provides an indication of the goodness of fit, listing the best-fit $\chi^2$ statistic and the corresponding degrees of freedom. The fit was overall good for all the sources, with the mean reduced $\chi^2$ of all the fits being $\chi^2_\nu=0.99$.

\subsection{Fe K$\alpha$ line}

\label{Sec:fe_line}

A detailed study of the iron's line spectral shape in the individual sources is outside the scope of this work. Therefore, we decided to fix the line width at 0.05 keV. There are several sources, however, for which the fit is significantly better when the width is left free to be minimised and the best-fit width is found to be larger than 0.05 keV even at a $3-\sigma$ level. We examined the fit's residuals when a narrow line is assumed for each of these sources and modified the fit as follows.

Following \cite{2015MNRAS.452.3266U}, we included three gaussian emission lines in our model for \textit{SWIFT J2209.4-4711}, all assumed to be narrow. The lines, which are centered at 6.4, 6.7, and 6.966 keV, model the emission from neutral Fe, Fe \texttt{XXV}, and Fe \texttt{XXVI}, respectively. The model with three lines provided an improved fit with $\Delta \chi^2=51$ in comparison to the model with only one narrow line.

The spectral fit of \textit{SWIFT J2304.8-0843} revealed the existence of a second emission line in its spectrum. Adding a second narrow line improved the fit by $\Delta \chi^2=17$ and the line was found to be centered at $E=6.93^{+0.08}_{-0.07}$ keV. The inclusion of a second line improved the fit significantly ($\Delta \chi^2=24$) for \textit{SWIFT J0433.0+0521}, as well. The existence of a line at $E\sim 6.9$ keV has already been found by \cite{2004MNRAS.354..839B}, who analysed an XMM observation of this source. The new fit results in the second line to be at $E=6.96^{+0.08}_{-0.10}$ keV. 

Moreover, the addition of a second line for \textit{SWIFT J1838.4-6524} led to a better fit with $\Delta \chi^2=7$. The new line was centered at $E=6.86 \pm 0.06$ keV. The same was true for two more sources. The fit was improved significantly, $\Delta \chi^2 = 10$ and 53, for the sources \textit{SWIFT J1347.4-6033} and \textit{J1836.9-5924}, respectively, when a second gaussian emission line was added to the model. The best-fit energy of the new line in the former source was $E= 6.88 \pm 0.06$ keV and $E=7.13 \pm 0.05$ keV for the latter one. For the five aforementioned sources the additional line may be explained as fluorescence from highly ionised, probably H-like, Fe atoms.

In addition to the sources with more than one emission lines, there were sources for which the fit resulted in a broad Fe K$\alpha$ line and no supplementary line was evident in the residuals. We fitted the corresponding spectra with the line width free to be minimised. There were nine such sources in total. Their names and the best-fit values of the line width are listed in Table \ref{tab:sources_feline}.

\begin{table}
	\centering 
	\caption{Sources with a broad emission line.} 
	\label{tab:sources_feline} 
	\begin{tabular}{ll} 
  \hline
	{Source Name}                & $\sigma_{Fe} (keV)$       \\
	                             &         \\
	\hline	
	
	SWIFT J0123.9-5846           & $0.21 \pm 0.03$          \\
	SWIFT J0244.8+6227           & $0.60 \pm 0.09$          \\
	SWIFT J0521.0-2522           & $0.75^{+0.13}_{-0.10}$   \\
	SWIFT J0925.0+5218           & $0.30^{+0.05}_{-0.04}$   \\
  SWIFT J1145.6-1819           & $0.40 \pm 0.08$          \\  
  SWIFT J1315.8+4420           & $0.33^{+0.06}_{-0.05}$   \\
  SWIFT J1349.3-3018           & $0.19 \pm 0.03$          \\
  SWIFT J1741.9-1211           & $0.40^{+0.12}_{-0.10}$   \\
  SWIFT J1835.0+3240           & $0.30 \pm 0.08$          \\
  
	\hline 
	\end{tabular}
\end{table} 


\subsection{Classification}
\label{Sec:class}

\begin{figure}
  \centering
  \includegraphics[width=\linewidth,height=0.7\linewidth, clip]{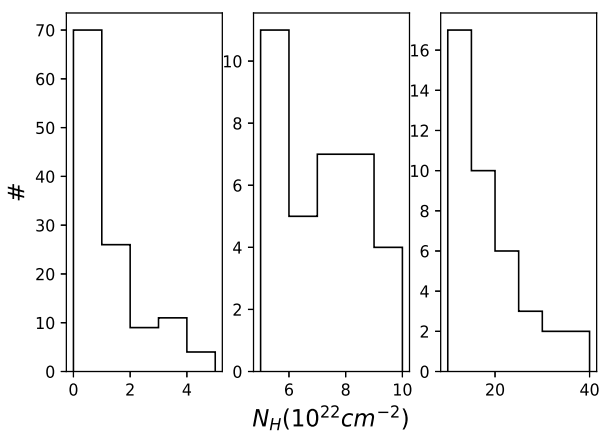}
  \caption[10]{Distribution of the best-fit $N_H$ in the different classes, UNOB (left), LOB (central), and MOB (right panel).}
  \label{fig:nh_hist}
\end{figure}

Before proceeding to the analysis of the spectral results, we found it useful to divide the sources into three groups based on their best-fit $N_H$ value. In this way, we were able to study the evolution of spectra as the absorption increases and to look for similarities and differences between the different groups. We chose to follow an $N_H$-based classification, because $N_H$ has a direct physical interpretation; it only depends on the amount of gas lying in our line of sight, and because differences in $N_H$ can be simply interpreted as differences in the source's inclination according to the unification model. In total, we defined three groups, the unobscured ($N_H< 5 \cdot 10^{22} \text{cm}^{-2}$, UNOB), the lightly obscured ($5 \cdot 10^{22} \text{cm}^{-2} < N_H < 10^{23} \text{cm}^{-2}$, LOB), and the mildly obscured ($N_H >  10^{23} \text{cm}^{-2}$, MOB) class. We chose an $N_H$ value of $5 \cdot 10^{22} \text{cm}^{-2}$ as the boundary between UNOB and LOB sources because $N_H$ values much smaller than this cannot be well constrained by fitting NuSTAR spectra, which start at 3 keV. The number of sources in each group versus their optical classification is given in Table \ref{tab:sources_class}. Figure \ref{fig:nh_hist} plots the distribution of $N_H$ in each class.

\begin{table}
	\centering 
	\caption{Number of sources in the $N_H$-based defined classes.} 
	\label{tab:sources_class} 
	\begin{tabular}{lrrrr} 
  \hline
	        &  Class 4       & Class 5 & \vline    &  Total   \\
	\hline	
 
  UNOB    &   92           &  28     & \vline    &  120     \\
  LOB     &   10           &  24     & \vline    &   34     \\
  MOB     &    5           &  48     & \vline    &   53     \\
  
	\hline 
	\end{tabular}
\end{table} 


\section{Results}
\label{sec:results}

\subsection{The photon index}

It is generally accepted that the X-ray AGN emission above 2 keV is mainly produced by Comptonisation of ultraviolet and optical photons in an optically thin region. Such an emission is well described by a power law. The power-law slope mainly depends on the accretion rate and on the corona's physical properties. As a result, variations in $\Gamma$ can be used  to probe variations close to the black hole.  AGN spectra have been found to exhibit a large range of indeces, from around 1.5 up to almost 2.5 with a mean value of around 1.8 \citep[e.g.][]{2017ApJS..233...17R}. 

\begin{figure}
  \centering
  \includegraphics[width=\linewidth,height=0.9\linewidth, clip]{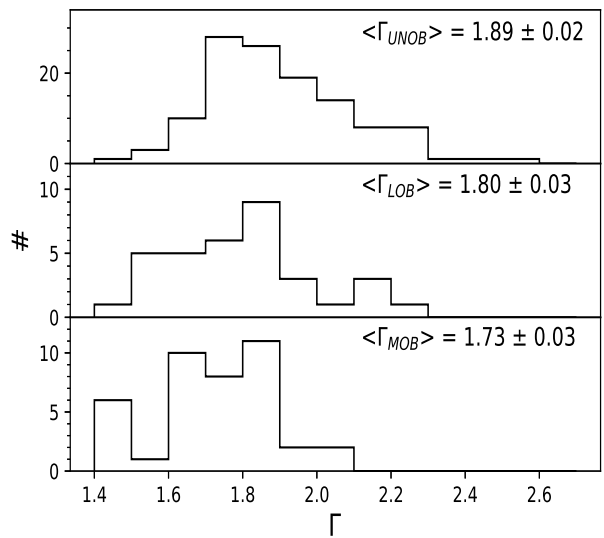}
  \caption[10]{Distribution of the best-fit $\Gamma$ in the UNOB (top), LOB (middle), and MOB (bottom panel) class.}
  \label{fig:gamma_hist}
\end{figure}

Figure \ref{fig:gamma_hist} plots the distribution of $\Gamma$, which seems to vary in each class. The main difference is that MOB sources occupy a shorter range of values than the other two classes, while sources with $\Gamma > 2.1$ are found only in less obscured, UNOB and LOB, groups. In addition, the photon index decreases with $N_H$, with the average value in each class being $1.89 \pm 0.02$, $1.80 \pm 0.03$, and $1.73 \pm 0.03$, respectively.

We evaluated the statistical significance of the difference in $\Gamma$ distributions using the Kolmogorov-Smirnov test. The LOB distribution was found to not be significantly different to neither the MOB nor the UNOB distribution. On the other hand, the difference between the distribution of UNOB and MOB sources was found to be significant, with a null hypothesis probability of $P_{null}=0.1\%$. The difference in the mean values between the two classes is $D\Gamma = 0.16 \pm 0.04$, and, hence, the two values are different even at a 3-$\sigma$ level. As discussed in Sect. \ref{sec:gamma_discrep}, the observed difference might be explained if a specific geometry is assumed for the X-ray source or if the corona is intrinsically different between higly absorbed and unabsorbed sources.

\subsection{The high-energy cutoff}

The high-energy cutoff may be viewed as a proxy for the temperature of the X-ray source. As a result, its knowledge provides information about the corona's dynamics and the physical processes that take place inside it \citep[e.g.][]{2015MNRAS.451.4375F}. Although a detailed discussion of the cutoff energy in individual sources is outside the scope of this work, some simple remarks could still be made.

The penultimate column of Table \ref{tab:best_fit} lists the best-fit values of the cutoff energy. For most of the sources it was not constrained and only a lower limit was derived. This was expected since the cutoff energy cannot be constrained when its real value is much higher than the energies probed by NuSTAR, or when the observation exposure and the source brightness are not high enough to allow its detection. More precisely, we constrained the cutoff for 55 sources and a lower limit was estimated for 152 sources. Taking into account only the well constrained values, the average high-energy cutoff for the total sample was found to be $E_c = 149 \pm 14 \text{ } \mathrm{keV}$. This value is consistent with the results of \cite{2019MNRAS.484.2735M} and \cite{2014ApJ...782L..25M}, who studied the hard X-ray spectrum of a sample of Type 1 AGN using NuSTAR and INTEGRAL, respectively. In the contrary, the calculated average is smaller than the value estimated by \cite{2017ApJS..233...17R} when they studied the BAT spectrum of Swift/BAT AGN taking into account the whole sample and it is also smaller than the value found by \cite{2014MNRAS.437.2845B} by fitting the luminosity function of local AGN. This apparent discrepancy is probably due to the fact that we only consider the constrained cutoffs in deriving the average value. 

\begin{figure}
  \centering
  \includegraphics[width=\linewidth,height=0.9\linewidth, clip]{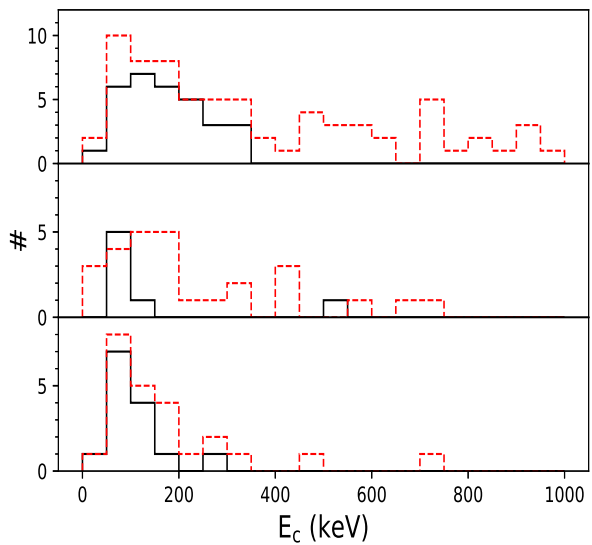}
  \caption[10]{Distribution of the best-fit $E_c$ in the UNOB (top), LOB (middle), and MOB (bottom panel) class. The solid black lines denote the distributions of the well constrained values and the red dashed lines denote the distribution of the estimated lower limits. }
  \label{fig:cutoff_hist}
\end{figure}

The distribution of cutoff in the three different classes is plotted in Fig. \ref{fig:cutoff_hist}. Using the Kolmogorov-Smirnov test and considering only the constrained values we found that no statistically significant difference was present between the different categories ($P_{null} > 1\%$). Furthermore, by estimating the corresponding Spearman's rank correlation, we looked for potential correlations between the high-energy cutoff and the other parameters. No statistically significant ($P_{null}$ was higer than 1\% in all cases) correlation was found between $E_c$ and $\Gamma$, $R$, and the X-ray luminosity of the source (Sect. \ref{sec:luminosities}).

\subsection{The reflection strength}

The reflection parameter, $R$, provides an estimation of how strong the reflected emission is with respect to the continuum power law. For a given inclination angle, $R=1$ corresponds to an isotropic X-ray source illuminating a slab disc. Constraining $R$ is important since its value provides information on the geometry of the source. In addition, the value of $R$ affects significantly the fraction of Compton thick sources estimated by cosmic X-ray background (CXB hereafter) population synthesis models. Since the two parameters are degenerated in reproducing the CXB spectrum, increasing $R$ decreases significantly the fraction of Compton thick AGN in the Universe.

Although several works have been conducted on determining $R$ in the various classes of AGN, its value is still debated. The results in the literature can often be contradictory and span a large range of values. Moreover, $R$ has been found to be correlated to other spectral or physical parameters of AGN. For instance, \cite{1999MNRAS.303L..11Z} found a positive correlation between $R$ and the spectral index, $\Gamma$. More recent studies have suggested a dependence of $R$ on the observed $N_H$ \citep{2011A&A...532A.102R} and X-ray luminosity \citep{2017ApJ...849...57D}. However, most of the results have been based on the analysis of average spectra. NuSTAR, being the first hard X-ray focusing telescope, provides a unique opportunity to study the evolution of reflection strength in individual sources to unprecedented detail.

\begin{figure}
  \centering
  \includegraphics[width=0.8\linewidth,height=0.9\linewidth, clip]{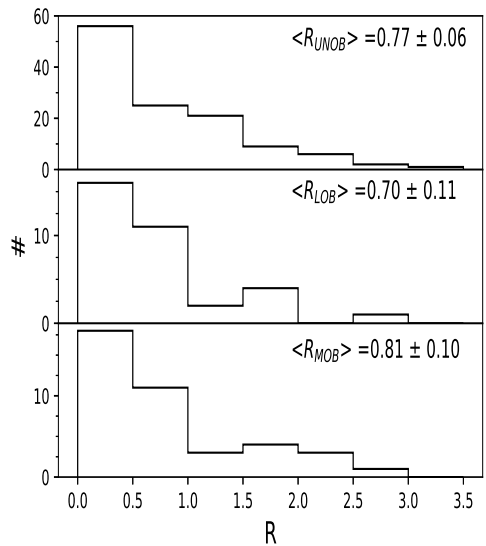}
  \caption[10]{Distribution of the best-fit $R$ in the UNOB (top), LOB (middle), and MOB (bottom panel) class.}
  \label{fig:refl_hist}
\end{figure}

Figure \ref{fig:refl_hist} plots the distribution of $R$ in the different classes. The three distributions look similar and no significant statistical difference was found. The average best-fit value for the total sample is $<R> = 0.78 \pm 0.05$. This is smaller than the usually assumed value of $R=1$ in CXB synthesis models, which is also expected for a simple disc reflection. However, our mean value is consistent within the errors with the estimations of previous studies \citep[e.g.][]{2011A&A...532A.102R, 2013ApJ...770L..37V}.

The mean value of reflection is $<R>=0.77 \pm 0.06$, $0.70 \pm 0.11$, and $0.81 \pm 0.10$ for the UNOB, LOB, and MOB sources, respectively. The mean values are fully consistent within the errors. This consistency is not easily explained within the simple unification model. To a first approximation, the reflection emission results from the scattering of the continuum power-law by the disc and torus surfaces. According to the unification model, the more obscured sources are observed at higher inclinations and, thus, the disc reflection is expected to be suppressed in the LOB and especially MOB sources due to geometric effects. If the torus is assumed to be similar in all AGN, then a decrease of $R$ with $N_H$ should be expected. The disagreement between this expectation and our results suggests that a more complicated mechanism is at work.

Strong reflection in highly absorbed sources has been firstly observed by \cite{2011A&A...532A.102R} using INTEGRAL observations and was later verified by \cite{2013ApJ...770L..37V} and \cite{2016A&A...590A..49E} using BAT data. This result could be explained if the covering factor of torus as observed by the central X-ray source is increasing with absorption. Hence, the total reflection would remain roughly constant despite the decrease of disc reflection. In a case of a clumpy torus, such an increase would correspond to an increase in the opening angle of the torus or to an increase of the clouds' filling factor.

\subsection{10-40 keV Luminosity}
\label{sec:luminosities}

Using the best-fit results we estimated the X-ray luminosity of each source from 10 to 40 keV. We calculated the intrinsic, meaning unabsorbed, luminosity and the corona luminosity, that is the luminosity predicted solely by the power-law emission component. The luminosity of MOB sources, which are the only sources with $N_H > 10^{23}\text{cm}^{-2}$, was also corrected for Thomson scattering because this is not taken into account by the used \textit{zphabs} model. The Thomson correction for sources with $N_H < 10^{23}\text{cm}^{-2}$ is smaller than 7\% having an insignificant effect to the results and was, thus, safely omitted for UNOB and LOB sources.

\begin{figure}
  \centering
  \includegraphics[width=\linewidth,height=0.7\linewidth, clip]{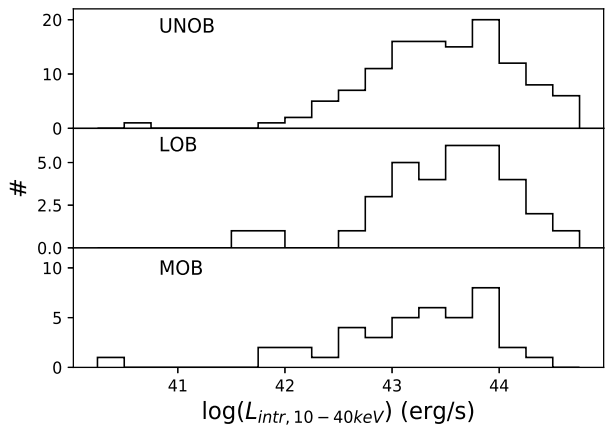}
  \caption[10]{Intrinsic 10-40 keV luminosity distribution in UNOB (top), LOB (middle), and MOB (bottom panel) sources.}
  \label{fig:lumintr_hist}
\end{figure}

\begin{figure}
  \centering
  \includegraphics[width=\linewidth,height=0.7\linewidth, clip]{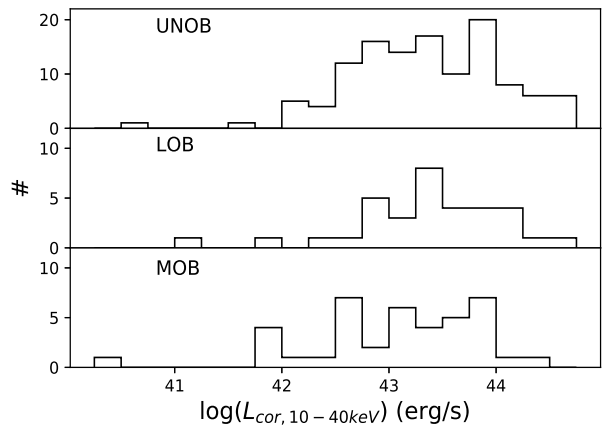}
  \caption[10]{Distribution of corona luminosity in the 10-40 keV energy range for UNOB (top), LOB (middle), and MOB (bottom panel) sources.}
  \label{fig:lumcor_hist}
\end{figure}

Figures \ref{fig:lumintr_hist} and \ref{fig:lumcor_hist} plot the distributions of intrinsic and corona luminosity, respectively. A visual examination of these figures reveals the similarity of the distributions in the different classes. The luminosities occupy a similar range of values in every class, with the majority of sources found in the range $42.5<\text{log}(L_{intr, 10-40 keV})<44$ in units of $\text{erg/s}$. There are only two sources, one in MOB and one in UNOB class, with an intrinsic luminosity $\text{log}(L_{intr, 10-40 keV})< 41.5 \text{ erg/s}$. While there is a tentative feature that highly luminous sources with $\text{log}(L_{intr, 10-40 keV})> 44 \text{ erg/s}$ are found mainly in a less obscured state, this result is not statistically significant. Moreover, the lack of any significant difference between the three classes was further supported by the results of the Kolmogorov-Smirnov test. However, this observed similarity is not surprising. The intrinsic source luminosity is expected to exhibit similar values regardless of the absorption state as the same central engine is assumed.

There have been numerous studies investigating the correlation of spectral parameters with X-ray luminosity. The existence of such correlations might shed light on the physical processes producing the different emission components. We, therefore, examined our results for the existence of any such correlation, as well.

The reflection strength was found to be slightly anti-correlated to the X-ray luminosity for the UNOB sources. A stronger correlation was found with the corona luminosity than with the intrinsic luminosity. The latter includes contribution from the reflected emission and is, consequently, expected to exhibit a weaker correlation. We calculated the Spearman's rank correlation coefficient between R and $L_{cor,10-40 keV}$ to be $\rho=-0.36$ with a chance probability of $P_{null}=5\cdot 10^{-5}$. However, it is likely that this correlation is observed only due to the selected sample. Sources with low power-law 10-40 keV luminosity are less likely to be detected by BAT, which is sensitive to photons with energy above 15 keV. It is then reasonable to assume that these sources would only be observed if they feature a strong reflection emission, which would render them brigther in total. A more detailed analysis taking into account the selection effects is needed before a robust conclusion on the reality of $R-L_{cor,10-40 keV}$ correlation can be made.

The reflection strength was not found to be correlated with the X-ray luminosity for either the LOB or the MOB sources. Moreover, no statistically significant ($P_{null} < 1\%$) correlation was found between the power-law index and the X-ray luminosity in either of the considered classes, consistent with previous studies \citep[e.g.][]{2009ApJ...690.1322W}.

\subsection{$R-\Gamma$ correlation}

In \cite{2019A&A...626A..40P}, we found that the reflection strength is positively correlated with the power-law index for unobscured sources. In the following, we examine the validity of this correlation in a much bigger sample and we discuss possible physical scenarios that could produce such a correlation.

\subsubsection{Unobscured sources}

\label{sec:unob_sources}

\begin{figure}
  \centering
  \includegraphics[width=\linewidth,height=0.7\linewidth, clip]{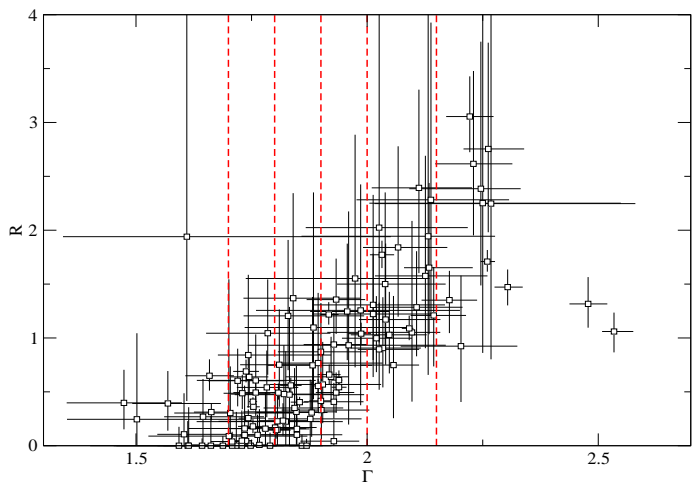}
  \caption[10]{$R$-$\Gamma$ correlation in UNOB sources. The vertical dashed lines indicate the boundaries between the different subgroups as defined in Sect. \ref{sec:unob_sources}.}
  \label{fig:refl_gamma_unob}
\end{figure}

The dependence of $R$ on $\Gamma$ for the UNOB class is ploted in Fig. \ref{fig:refl_gamma_unob}. The two parameters are clearly correlated. However, they are also degenerated, meaning that they are intrinsically correlated in the model. Consequently, a robust conclusion cannot be made by a simple visual examination of the scatter plot. We followed different approaches in order to verify that the observed correlation is real.

\begin{figure}
  \centering
  \includegraphics[width=\linewidth,height=0.7\linewidth, clip]{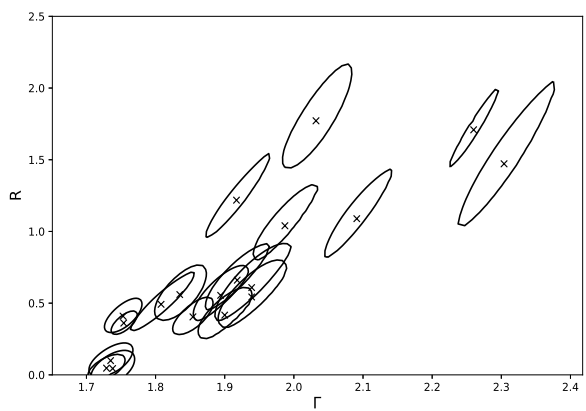}
  \caption[10]{Observed $R$-$\Gamma$ correlation for the UNOB sources with well constrained $\Gamma$ (see text for details). The error contours correspond to a 99\% confidence interval ($\Delta\chi^2=9.21$).}
  \label{fig:refl_gamma_unob_cont}
\end{figure}

Firstly, we restricted our sample to the sources with high quality data. More precisely, we studied only the sources for which $\Gamma$ is well constrained, with a 1-$\sigma$ relative error of $\frac{\Delta \Gamma}{\Gamma} < 0.0135$. There were 19 such sources. For each one of them, we calculated their error contour using the \textit{steppar} command in XSPEC. The results are plotted in Fig. \ref{fig:refl_gamma_unob_cont}. The plotted contours correspond to a confidence interval of 99\% ($\Delta\chi^2=9.21$). The general trend of $R$ increasing on average with $\Gamma$ is well observed. It is evident that the best-fit values are not consistent within the errors, and their differences cannot be explained by the observed degeneracy. 

\begin{figure}
  \centering
  \includegraphics[width=\linewidth,height=0.7\linewidth, clip]{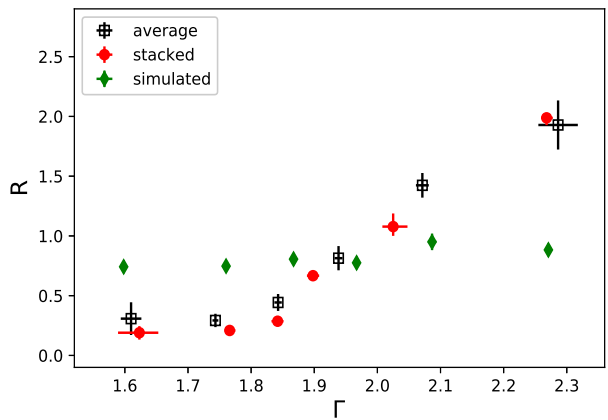}
  \caption[10]{Observed $R$-$\Gamma$ correlation for the subgroups of UNOB class. The red filled circles correspond to the best-fit values of the stacked spectra, while the black open squares to the average values of the individual best fits. The green diamonds correspond to the mean values retrieved for the simulated data. The errors are not distinguishable because of their small value.}
  \label{fig:refl_gamma_unob_stack}
\end{figure}

Furthermore, we produced simulated spectra to assess how the model's degeneracy affects the estimation of the two parameters. Using the NuSTAR responses, we simulated \footnote{The simulated spectra were produced using the \textit{fakeit} command in XSPEC.} the spectra expected to be observed from sources with emission given by the \textit{pexrav} model. Spectra were produced for the same value of reflection strength, equal to the average value of UNOB objects, $R=<R_{UNOB}>=0.77$ and for six different values of spectral indeces, $\Gamma= 1.6, 1.75, 1.85, 1.95, 2.07, \text{ and } 2.25$. These values correspond roughly to the mean values of each subgroup defined below in Sect. \ref{sec:unob_sources}. For each $\Gamma$, 200 simulated spectra were produced assuming a power-law normalisation equal to the mean value of each subgroup and an exposure time of 25 ks. We then fitted the simulated data with the \textit{pexrav} model and calculated the mean $R$ and $\Gamma$ for each group of simulated spectra in order to evaluate whether the observed trend could be an artifact of the quality of our data and the model's degeneracy. The estimated mean values are plotted as green diamonds in Fig. \ref{fig:refl_gamma_unob_stack}. It is evident that the model's degeneracy does not reproduce an average increase of reflection with the power-law slope. The $R$ values are consistent to the mean value within the errors. We, therefore, concluded that the observed correlation is real. The reality of this correlation for a smaller sample of sources and at lower energies was also studied in detail and confirmed by \cite{2003MNRAS.342..355Z}.

Having confirmed the reality of the observed correlation, it should be noted that this is not a 1:1 correlation. There are two main reasons why scattering around a single line correlation is expected. Firstly, statistical errors might result in a scatter around the real correlation. Secondly, and more importantly, a prominent scatter around a 1:1 correlation is expected because of the physical differences in different AGN. AGN are really dynamic systems occupying a range of black hole mass, accretion rate, black holes spin, and several other physical parameters that would define the system uniquely. It is then expected that, for example, two sources with the same reflection strength could have different power-law slopes as a result of the intrinsic physical differences between them. Therefore, the observed trend suggests that only on average sources with higher reflection strength would also feature steeper spectra for the continuum emission.

In order to improve the statistics, we did not consider the best-fit values of the individual sources in the following analysis. Instead, we divided the UNOB sources into six groups based on their best-fit $\Gamma$ value and estimated the mean value of the individual fits for both $R$ and $\Gamma$ in each group. The boundaries of each group are denoted by vertical red dashed lines in Fig. \ref{fig:refl_gamma_unob}. In addition, we calculated the stacked spectrum of each subgroup. In deriving the stacked spectra we renormalised each individual spectrum before adding with respect to its ARF file in order to account for differences in the source's size. The stacked spectra were calculated using the \textit{mathpha} tool and were then binned so that each bin contains at least 100 source counts.

Figure \ref{fig:refl_gamma_unob_stack} plots the dependence of $R$ on $\Gamma$ after grouping the sources. The black open squares denote the mean values of the individual fits, taking into account all the UNOB sources, and the red filled circles denote the best-fit values of the stacked spectra. The two sets predict a similar correlation trend, providing extra evidence that this is a real correlation. The explanation of this trend by physically motivated models is discussed in Sect. \ref{sec:phys_interp}.

\subsubsection{The special case of SWIFT J0947.6-3057}

\begin{figure}
  \centering
  \includegraphics[width=\linewidth,height=0.8\linewidth, clip]{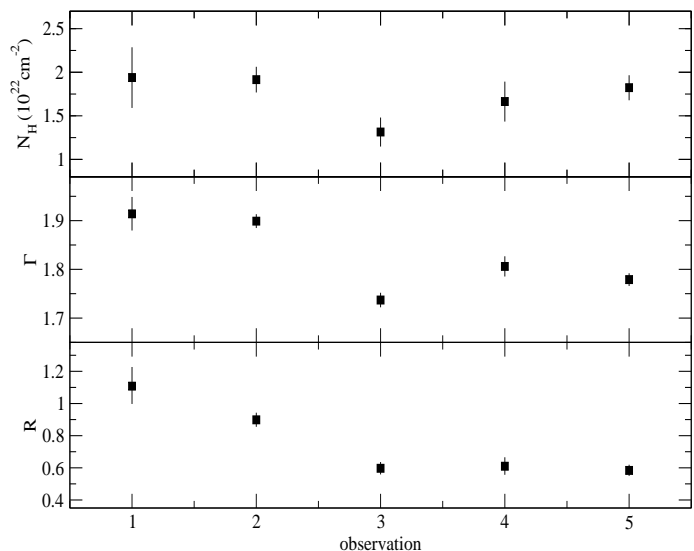}
  \caption[10]{Best-fit absorption column density (top), spectral index (middle), and reflection strength (bottom panel) for SWIFT J0947.6-3057. The x axis corresponds to an increasing number of observation.}
  \label{fig:mcg_fit}
\end{figure}

\textit{SWIFT J0947.6-3057} was observed five times with NuSTAR from July 2012 until March 2015. We fitted all the five spectra of the source with the baseline model (Sect. \ref{Sec:base_model}). The best-fit results are listed in Table \ref{tab:best_fit} and plotted in Fig. \ref{fig:mcg_fit}. \textit{SWIFT J0947.6-3057} is the only object of our sample that exhibits a strong variation in the power-law slope and reflection strength, while its absorption level remains roughly constant. The lack of variability for $N_H$ is supported by a $\chi^2$ test ($P_{null}=10\%$), unlike for $\Gamma$ and $R$. This feature let us study the variations of $R$ with respect to $\Gamma$ variations for an individual source.

\begin{figure}
  \centering
  \includegraphics[width=\linewidth,height=0.7\linewidth, clip]{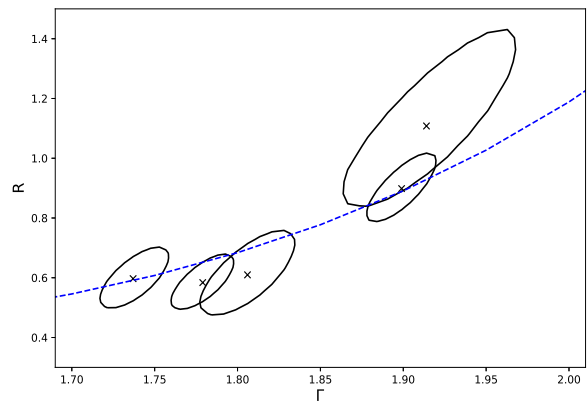}
  \caption[10]{$R$-$\Gamma$ correlation in SWIFT J0947.6-3057. The x symbols indicate the best-fit values, while the error contours correspond to a 99\% confidence level ($\Delta\chi^2=9.21$). The dashed blue line indicates the correlation expected for a moving corona (Sec. \ref{sec:phys_interp}). It is similar to the one plotted in Fig. \ref{fig:refl_gamma_unob_stack2}, but a constant of $R=0.25$ has been added to it.}
  \label{fig:mcg_cont}
\end{figure}

Figure \ref{fig:mcg_cont} plots the best-fit $R$ as a function of $\Gamma$. The crosses indicate the best-fit values and the ellipses denote the best-fit contour of the two parameters at a 99\% confidence interval, as calculated using the \textit{steppar} command in XSPEC. The observed correlation cannot be explained due to the statistical errors or the degeneracy of the two parameters. We concluded that this is a real correlation, intrinsic in the emission of the source.

\cite{2017ApJ...836....2Z} have suggested that an apparent correlation between the reflection strength and the photon index might be observed for a single source even when the reflected emission remains constant. The emission from a distant reflector smooths out the variability of the primary continuum and, consequently, two observations might exhibit the same amount of reflected emission but a different continuum power law. If the continuum is steeper in one observation, then the reflection strength in this observation will also be larger in order to match the observed reflected flux. Therefore, the observed correlation might be an artifact of the model fitting. However, this is not the case for \textit{SWIFT J0947.6-3057}. Using our best-fit results, we confirmed that the observations of larger $R$ feature higher flux of reflected emission, as well. In other words, the increase in $\Gamma$ was found to correspond to an increase in the flux of the reflected emission, excluding the possibility that the observed correlation is an artifact of the fit. This is further supported by the results of \cite{2014ApJ...789...56Z}. These authors found that the Compton hump emission lags behind the continuum power-law by $\sim$ 1 ks, strongly favouring a disc reflection for large part of this emission, instead of a distant reflector; while, the existence of an $R-\Gamma$ correlation requires the disc to be the main reflector, as well (Sect. \ref{sec:phys_interp}).

An $R$-$\Gamma$ correlation has been observed before for a group of Seyfert galaxies \citep[e.g.][]{1999MNRAS.303L..11Z}. However, to the best of our knowledge, this is the first time that a positive correlation is found for an individual AGN with such a high statistical significance. The importance of this result lies, among other things, on the fact that this would provide further constrains on the proposed explanations of the $R$-$\Gamma$ dependence, if the same driving mechanism is assumed. This mechanism should be able to account for both variations in the local environment of an AGN and variations between different sources.

As discussed in Sect. \ref{sec:phys_interp}, the most promising model to explain the observed correlation is that of an X-ray source moving with respect to the disc. The dashed blue line in Fig. \ref{fig:mcg_cont} plots the predicted relation of this model. No fitting of this line has been applied to the data, but a constant value of $R=0.25$ has been added to the predicted values in order for them to match the observed ones. This amount of reflection might be attributed to the reflection originating from a second scattering surface, like the surrounding torus. \textit{SWIFT J0947.6-3057} is a Seyfert 2 galaxy and a torus contribution to the reflection emission is not surprising. The predicted line is fully consistent with the observed values within the errors.

\subsubsection{Obscured sources}
\label{sec:obs_sources}

We have shown previously that an $R$-$\Gamma$ correlation is not observed for the obscured sources, pointing to a different behaviour of the reflection emission between obscured and unobscured AGN \citep{2019A&A...626A..40P}. The lack of an $R$-$\Gamma$ correlation is, however, not unexpected in this case. In general, the obscured sources are assumed to be observed at higher inclination angles and, thus, less disc emission will be observed. Therefore, a subtler correlation is expected to be observed, if any.

\begin{figure}
  \centering
  \includegraphics[width=\linewidth,height=0.7\linewidth, clip]{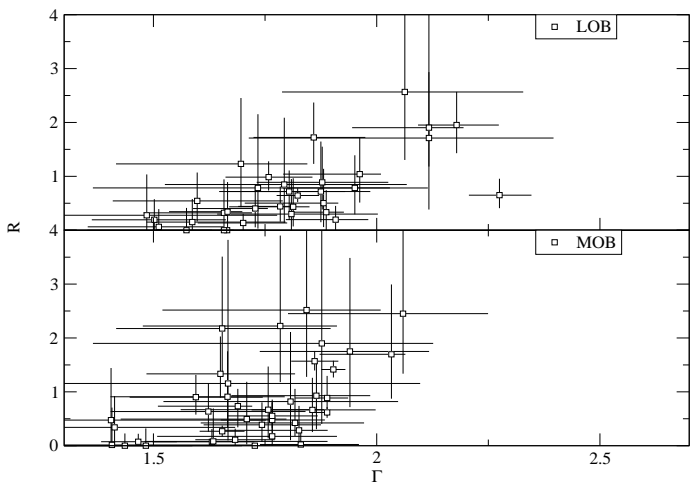}
  \caption[10]{Best-fit reflection strength against power-law slope for the LOB (upper) and MOB (lower panel) sources. The axes are of the same scale as in Fig. \ref{fig:refl_gamma_unob}. }
  \label{fig:refl_gamma_obs}
\end{figure}

Figure \ref{fig:refl_gamma_obs} plots the best-fit $R$ versus $\Gamma$ for the obscured sources, that is for both LOB and MOB objects. The axes are of the same scale as in Fig. \ref{fig:refl_gamma_unob} to allow for a direct comparison. The two parameters seem to be only slightly correlated, but a visual examination of Fig. \ref{fig:refl_gamma_obs} does not allow for a robust conclusion on the existence of a positive correlation. Interestingly enough though, the sources in LOB class exhibit slightly stronger evidence of a correlation.

To evaluate the existence of a real correlation, the degeneracy of the two model parameters has again to be taken into account. Due to the larger errors and the smaller amount of objects in the MOB and LOB classes, there are not sufficiently enough sources with well constrained $\Gamma$ to allow studying a subsample of high quality data, similarly to the analysis followed in Sect. \ref{sec:unob_sources}. Instead, we produced simulated data to examine the reality of the observed trend.

We generated two groups of simulated spectra. Initially, 200 spectra were generated assuming a \textit{pexrav} model for the source emission with a reflection, spectral index and power-law normalisation equal to the average values of the best-fit results for LOB sources. An exposure time of 25 ks was assumed and the NuSTAR responses were utilised. The simulated spectra were then fitted using the \textit{pexrav} model and the resulted $R$-$\Gamma$ values were fitted by a straight line. In this way, we were able to measure and describe in an analytical form the apparent correlation produced by the model's degeneracy. The same procedure was followed to produce another 200 fake spectra, using this time the best fit results of the MOB sources. Eventually, we had parametrised the degeneracy expected for the two classes by two straight lines.

In the following, we tested if the $R$-$\Gamma$ trend observed in the obscured sources could be produced by the model's degeneracy. To that extent, we assumed that all the sources of each class have the same $R$ value, equal to the corresponding average value, and the observed difference between this value and the measured one is the result of the degeneracy. Having parametrised this degeneracy by a straight line, we were able to project every measured point ($R$,$\Gamma$) onto the line $R=<R>$ and as a result, estimate a new value for the power-law slope, $\Gamma_{d}$. This new value corresponds to the real value of the power-law emission under the assumption of a constant $R$. If this new value deviates significantly from the initially calculated best-fit $\Gamma$, then the hypothesis of constant $R$ is probably wrong. If, instead, the two values are consistent within the errors, the hypothesis cannot be rejected. 

\begin{figure}
  \centering
  \includegraphics[width=\linewidth,height=0.7\linewidth, clip]{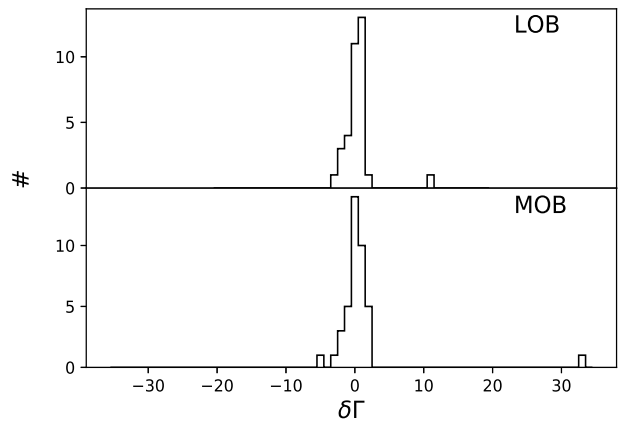}
  \caption[10]{Histogram of the quantity $\delta \Gamma$, as defined in Sect. \ref{sec:obs_sources} for the two classes of obscured objects.}
  \label{fig:dgamma_obs_hist1}
\end{figure}

\begin{figure}
  \centering
  \includegraphics[width=\linewidth,height=0.7\linewidth, clip]{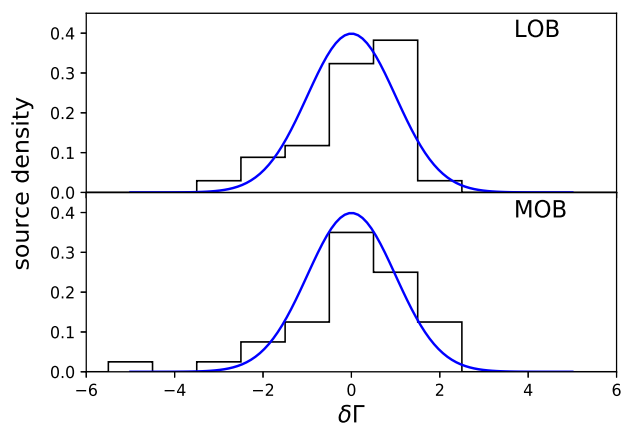}
  \caption[10]{$\delta \Gamma$ normalised histogram. One LOB (upper panel) and one MOB (lower panel) source lie outside the plotting range. The blue solid line is the same in both panels and corresponds to the standard normal distribution.}
  \label{fig:dgamma_obs_hist2}
\end{figure}

Figure \ref{fig:dgamma_obs_hist1} plots the histogram of this deviation quantified as $\delta \Gamma = \frac{\Gamma - \Gamma_{d}}{\Delta \Gamma}$, where $\Delta \Gamma$ is the error of the best-fit value $\Gamma$. The upper and lower panel correspond to the LOB  and MOB sources, respectively. Figure \ref{fig:dgamma_obs_hist2} is a zoomed version of Fig. \ref{fig:dgamma_obs_hist1}. It should be noted that the histogram in Fig. \ref{fig:dgamma_obs_hist2} has been renormalised, meaning that the source density, instead of the number of sources, is plotted in this figure. The blue solid line denotes the normal distribution of zero mean and unity standard deviation. No fit has been applied to the histogram.

Only one LOB and one MOB source exhibit a strong deviation between the best-fit $\Gamma$ and the one predicted under the assumption of constant reflection. The deviation in the remaining sources seem to be consistent with zero within the errors. This implies that only for a few sources a different value of reflection is statistically required. Taking into account the simplicity of our initial hypothesis and the diversity of AGN, our main conclusion is that the quality of the data does not allow for a statistically robust detection of an $R$-$\Gamma$ correlation in the obscured sources. The observed trend can be well explained as the result of the model's degeneracy. Instead, there seems to be a correlation between $R$ and $N_H$ in these sources. A detailed study of this correlation will be presented in an upcoming paper.

\section{Discussion}
\label{sec:discuss}

\subsection{$\Gamma$ discrepancy}
\label{sec:gamma_discrep}

The applied analysis revealed a difference in the power-law slopes between unabsorbed and heavily absorbed sources. While there was no prominent disagreement between the $\Gamma$ distributions of LOB and the other two classes, MOB sources were found to exhibit harder spectra than UNOB sources. A Kolmogorov-Smirnov test confirmed the statistical significance of this difference ($P_{null}=0.1\%$), with the mean $\Gamma$ values of the two classes differing by $D\Gamma = 0.16 \pm 0.04$.

We checked whether the observed difference could be a selection effect of the used sample. More precisely, we examined if the MOB sources are intrinsically less luminous than the UNOB or if the UNOB objects with large best-fit $\Gamma$ are less luminous in comparison to UNOB sources with small $\Gamma$. Any of these two cases, might lead to high-$\Gamma$ MOB sources not be detected by BAT due to their high absorption. In Sect. \ref{sec:luminosities}, we estimated the corona luminosity of all the sources in the energy range 10-40 keV. This energy range is well probed by NuSTAR for most of our sources and highly overlaps with the energy range of BAT. We found no significant differences between the luminosity of MOB and UNOB sources or between the luminosity of sources with different $\Gamma$. As a result, we concluded that the difference in the $\Gamma$ distributions is real and not a selection effect.

Since the AGN X-ray continuum is the result of Comptonisation, a harder spectrum, meaning smaller $\Gamma$, is expected from an optically thicker Comptonising region. Assuming that UNOB and MOB AGN feature no intrinsic differences in the physical properties of the X-ray source, a higher optical depth in MOB sources indicates that soft photons spend on average more time within the X-ray source in absorbed sources than in unabsorbed ones. This might be achieved if a slab geometry is assumed for the corona. According to the unification model, highly absorbed sources are observed on average at higher inclination angles than unabsorbed sources. If the corona is homogeneous and of a slab shape, then the photons emitted closer to edge-on had on average longer paths to travel within the corona, and thus correspond to larger optical depth. 

Using the \textit{compps} model \citep{1996ApJ...470..249P} in XSPEC, we confirmed that a slab corona would produce a harder spectrum when observed closer to edge-on. The difference between the edge-on and face-on $\Gamma$ was estimated to be about 0.07 for vertical optical depths of $\tau \sim 1$. This value is somewhat smaller than the observed discrepancy ($D\Gamma = 0.16 \pm 0.04$), but consistent to it within $\sim$2.3 $\sigma$. If such a high difference in $\Gamma$ is indeed verified with higher significance, inclination effects might only be partly contributing to this discrepancy.

The observed disagreement might also be explained if different parts of the X-ray source are observed in the different classes. For example, it is reasonable to assume that the X-ray source is vertically extended above the disc. Sources observed close to face-on will then be dominated by the emission of the source's top part, while sources of higher inclination will feature emission from both upper and lower regions of the X-ray source. If a specific vertical structure of the X-ray source is assumed, which could, for instance, be variations in its temperature profile, one may expect a different spectral shape for the emission originating from different layers of the source. Although tempting, a detailed estimation of the expected discrepancy in this case would require the knowledge of the X-ray position source and its structure. Such an analysis is outside the scope of this work.

\subsection{Physical interpretation of the $R$-$\Gamma$ correlation}
\label{sec:phys_interp}

The results of the spectral analysis suggested a positive correlation between the reflection strength and the spectral index for the UNOB sources. The $R$-$\Gamma$ correlation may provide information about the inner geometry of AGN. The power-law index describes the shape of a Comptonisation spectrum and it, thus, depends on the flux and spectrum of seed disc photons and on the physical state of the corona. On the other hand, the reflection strength depends on the X-ray primary continuum emission and on the characteristics of the scattering surface that produces the reflection emission. It is then evident that the two parameters can be correlated only if the disc is the primary reflector in the UNOB sources. This result is further supported by the detection of high-frequency soft or Fe K$\alpha$ lags in the emission of several UNOB sources. These lags, usually of the order of ten to hundreds of seconds, correspond to time difference between variations in the X-ray continuum and variations in the soft X-rays or Fe line, with the continuum driving the variability. It has been argued that the observed lags can be explained if the delayed emission is produced due to scattering of the inner disc surface \citep[e.g.][]{2009Natur.459..540F}. In total, at least seven sources have been found to exhibit a soft lag and seven sources exhibit a Fe lag \citep[][and references therein]{2013MNRAS.431.2441D, 2016MNRAS.462..511K}.

It should also be noted that $R$ exhibits values importantly higher than 1 for $\Gamma>2.0$. If a thin disc is the sole reflector of the primary emission, such high values of reflection strength are not easily explained and an anisotropy of the X-ray source or some relativistic effects have to be assumed.

Moreover, the exact shape of the correlation provides important constrains on the geometry or the physical interplay between the disc and the corona. \cite{1999MNRAS.303L..11Z} were the first to observe a positive correlation between the reflection strength and the power-law index in Seyfert galaxies. They concluded that such a correlation could be explained if the corona is moving with respect to the disc, a model derived by \cite{1999ApJ...510L.123B}. In this model, magnetic flares are produced above the disc with a slightly relativistic velocity. When the flare is moving away from the disc, relativistic aberration reduces the amount of photons emitted towards the disc and subsequently the reflected emission, which also leads to a harder spectrum. The opposite trend is expected when the flare is moving towards the disc.

\begin{figure}
  \centering
  \includegraphics[width=\linewidth,height=0.7\linewidth, clip]{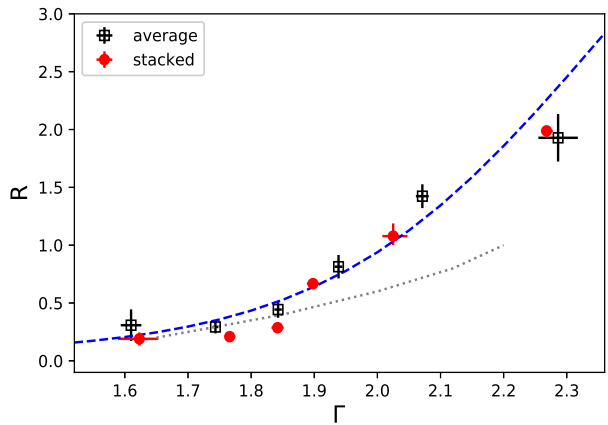}
  \caption[10]{Observed $R$-$\Gamma$ correlation for the subgroups of UNOB class. The red filled circles correspond to the best-fit values of the stacked spectra, while the black open squares to the average values of the individual best fits. The errors are not distinguishable because of their small value. The dashed blue line indicates the correlation expected for a moving corona with respect to the disc \citep{1999ApJ...510L.123B} and the dotted gray line indicates the correlation predicted for a variable covering disc factor.}
  \label{fig:refl_gamma_unob_stack2}
\end{figure}

Figure \ref{fig:refl_gamma_unob_stack2} plots the observed $R-\Gamma$ correlation, exactly as in Fig. \ref{fig:refl_gamma_unob_stack}. Using the equations of \cite{1999ApJ...510L.123B}, we calculated the correlation predicted for a moving X-ray source, which is plotted as a blue dashed line in Fig. \ref{fig:refl_gamma_unob_stack2}. The blue line was produced assuming a disc albedo of $\alpha=0.15$, a system inclination of $\text{cos } i=0.45$, and a value of 0.55 for the geometry-dependent parameter $\mu_s$, which is defined in the aforementioned paper. No fitting was performed of this line to the data. The predicted correlation seems to explain the observed trend well.

The model requires a velocity range of $-0.3<\beta<0.6$ for the observed range of $\Gamma$ and $R$ to be reproduced, where $\beta$ is the source velocity in units of light speed and the minus sign indicates move towards the disc. The deduced velocities are only mildly relativistic. Although it is difficult to conceive a mechanism to accelerate the X-ray source both outwards and towards the disc, this possibility cannot be excluded. For example, \cite{1999ApJ...510L.123B} proposed a radiatively driven acceleration mechanism.

This model is able to explain the observed correlation for \textit{SWIFT J0947.6-3057}, as well. The variation in $\Gamma$ for this object suggests a variation in the source velocity from $\sim 0.4$ to $\sim 0.16$ in units of light speed, which is not unreasonably high. Since variations in the source velocity are expected to depend on variations of the inner disc, where the source is formed, the observed velocity variability could be achieved within the spectral variability timescale of \textit{SWIFT J0947.6-3057}. 

The predicted correlation for this object is plotted in Fig. \ref{fig:mcg_cont} as a blue dashed line. A constant value of $R=0.25$ has been added to this line in comparison to the line of Fig. \ref{fig:refl_gamma_unob_stack2} in order to match the observed values. This extra emission might be the result of additional reflection from surrounding material. The predicted correlation reproduces well the observed trend within the errors.

A positive $R$-$\Gamma$ correlation might also be produced if the disc covering factor as observed from the primary X-ray source is variable. It has been suggested that the continuum X-ray emission might originate from a hot optically thin flow in the inner part of the disc. In this configuration, matter accretes in the form of a Shakura-Sunyaev disc \citep{1973A&A....24..337S} until a specific radius, after which the disc is truncated. At radii smaller than the truncation radius, a complicated accretion might be in place, with hot optically thin accretion flow and cold clumps of an optically thick accretion disc coexisting. Recently, \cite{2018A&A...614A..79P} studied the spectral emission for such a configuration. They found that for a given ionisation level the power-law index and the disc covering factor, $f_c$, are positively correlated (Fig. 5 in their paper). Since for a given inclination angle the reflection strength corresponds to the reflector's covering factor, the above result predicts a correlation between $\Gamma$ and $R$, as well. Using their results and for $R=f_c$, we calculated the predicted correlation, which is plotted as a gray dotted line in Fig. \ref{fig:refl_gamma_unob_stack2}. The model predicts a weaker correlation than the observed one and is unable to probe all the observed values. The power-law slope could reach values above 2.2 if disc dissipation is assumed. However, the reflection strength might reach values above unity only if some relativistic beaming is assumed or under the assumption of an anisotropic source. 

The existence of the correlation for a single source renders this model even less plausible. Variations of the disc covering factor in one source are expected to happen at timescales similar to or larger than the viscous timescale, which is given in seconds by \citep{2002apa..book.....F}:

\begin{eqnarray}
    t_{visc} \sim 3 \times 10^5 \alpha^{-4/5} \dot{M}_{16}^{-3/10}M_1^{1/4}R_{10}^{5/4},
\end{eqnarray}

\noindent where $\alpha$ is the parameter describing the disc viscosity, $\dot{M}_{16}$ is the mass accretion rate in units of $10^{16} g/s$, $M_1$ is the black hole mass in units of solar masses and $R_{10} = R/(10^{10} \text{cm})$. For an accretion rate $\dot{M} = 0.05 \dot{m}_{Edd}$ ($\dot{m}_{Edd}$ is the Eddington accretion rate), a black hole mass of $M_{BH}=2\cdot10^6$ solar masses \citep{2007ApJ...660.1072W}, and assuming a viscosity parameter of $\alpha=0.1$, the viscous timescale at $R=50 R_g$ ($R_g$ is the gravitational radius) for \textit{SWIFT J0947.6-3057} is of the order of 150 years. However, $R$ and $\Gamma$ seem to vary significantly within 2 years. As a result, this variation cannot be explained by variations in the disc geometry.

Another possible way for the reflection strength to vary would be if the position of the corona with respect to the black hole is changing. Assuming a lamp post geometry, the reflection strength will increase when the corona is located closer to the black hole because gravitational effects would lead to the illumination of disc by a larger flux than the one observed along our line of sight. However, it is not a straightforward exercise to imagine why the power-law slope would then also increase. One possible solution would be if the seed disc emission observed by the corona is higher, leading to faster cooling and consequently to a softer spectum. At first approximation, however, and ignoring relativistic effects, the disc emission intercepted by a region closer to the black hole would be smaller because the disc is now viewed under a smaller solid angle. A detailed fully relativistic treatment is needed to examine whether a variation in the corona's position could reproduce the observed trend. 

Furthermore, this explanation does not seem to be able to account for the correlation in \textit{SWIFT J0947.6-3057}. The corona flux for this source in 10-40 keV, that is the flux predicted by the power-law model component at this energy range, remains constant over the observation period. If the corona is located closer to the black hole for some of the observations, the corona flux would be expected to decrease, unless a fine tuning between the corona's position and the power-law normalisation is assumed.

Finally, it has been suggested that a varying ionisation level for the disc surface might produce a positive $R$-$\Gamma$ correlation. \cite{2001ApJ...546..419D} showed that using a single-ionisation model to fit a reflection spectrum originating from a disc with multiple ionisation layers can produce an apparent correlation. However, the estimated correlation (Fig. 2 in the aferomentioned paper) is much weaker than the observed one. The authors also mentioned that a model calculating self-consistently the expected $\Gamma$ might increase the estimated correlation. It is not easy to conceive why this would be the case though. Instead, it would be expected that variations in the ionised disc surface will result in an anti-correlation between $R$ and $\Gamma$. The two authors suggested that an increase in the optical depth of a highly ionised disc surface would result in smaller $\Gamma$. At the same time though, increasing the optical depth would increase the albedo of the disc's surface, which will now behave as a mirror for the incident X-rays. This should result in a significant increase of the reflection. As a result, the two parameters would then be anti-correlated.

\section{Conclusions}
\label{sec:conclude}

In this work, we studied the NuSTAR spectra of a large sample of nearby Seyfert galaxies. Thanks to the sensitivity of NuSTAR, we studied the reflection hump of these sources in unprecedented detail. Our main findings could be summarised as follows.

The reflection strength was found to be positively correlated with the power-law slope in unabsorbed sources. Such a correlation strongly favours the disc to be the main reflector of the X-ray continuum emission. The same correlation was, also, found for the case of the individual source \textit{SWIFT J0947.6-3057}. Although different mechanisms could, in principle, result in a positive correlation between $R$ and $\Gamma$, not all of them were able to reproduce the exact shape of the observed correlation or to explain the large range of observed $R$ values. The most promising explanation is that electrons in the corona are moving with respect to the disc at a moderately relativistic velocity. This model can also explain the existence of the correlation in a single source.

On the other hand, no $R-\Gamma$ correlation was detected in the case of absorbed sources, although both absorbed and unabsorbed sources feature similar levels of reflection. This result might indicate a different origin for the reflection emission of absorbed AGN.

Furthermore, both absorbed and unabsorbed sources were found to exhibit similar luminosity values. Interestingly, the luminosity of individual AGN didn't seem to be driven by the shape of the X-ray spectrum.

Finally, the heavily absorbed sources featured harder spectra than unabsorbed sources. The observed difference in mean $\Gamma$ between the two groups of objects could be explained if the corona has a slab geometry.

\bibliographystyle{aa} 
\bibliography{arxiv_draft} 

\begin{table*}
	\centering 
	\caption{Excluded sources} 
	\label{tab:sources_exclud} 
\begin{tabular}{llc} 
  \hline
	{Source Name}                &  Counter part             & Exclusion reason       \\
	                             &                           &         \\
	\hline				
	SWIFT J0006.2+2012           & Mrk 335                           &  V     \\
	SWIFT J0009.4-0037           & 2MASX J00091156-0036551           &  F     \\
	SWIFT J0025.8+6818           & 2MASX J00253292+6821442           &  RD    \\
	SWIFT J0101.5-0308           & 2MASX J01012440-0308399           &  F     \\
	SWIFT J0107.7-1137B          & 2MASX J01073963-1139117           &  F     \\
	SWIFT J0111.4-3808           & NGC 424                           &  RD    \\
\hline 
	\end{tabular}
\end{table*} 

\addtocounter{table}{-1}		

\begin{table*}
	\centering 
	\caption{Continued.} 
	\label{tab:sources_exclud} 
	\begin{tabular}{llc} 
  \hline
	{Source Name}                &  Counter part             & Exclusion reason       \\
	                             &                           &         \\
	\hline				
	SWIFT J1309.2+1139           & NGC 4992                          &  RD    \\
	SWIFT J1322.2-1641           & MCG -03-34-064                    &  RD    \\
	SWIFT J1333.5-3401           & ESO 383- G 018                    &  RD    \\	
	SWIFT J1355.9+1822           & Mrk 463                           &  RD    \\
	SWIFT J1416.9-4640           & IGR J14175-4641                   &  RD    \\
	SWIFT J1424.2+2435           & NGC 5610                          &  F     \\
\hline 
	\end{tabular}
	\tablefoot{The first two columns list the Swift BAT name of the source and the name of its counter part. The last column lists the reason for excluding the source from the current analysis; F: low signal-to-noise ratio; L: LINER; RD: reflection dominated spectrum; V: high spectral variability; PC: source in Perseus cluster.}
\end{table*}

\begin{table*}
	\centering
	\caption{Source sample used in this work.}
	\label{tab:sources_info}
	\begin{tabular}{llrrrcr} 
		\hline
		{Source Name}                & Counter part      &  RA                       &Dec             &z      &BAT Class &$N_H$ class    \\
		                             &                   &                           &                &       &          &               \\
		\hline
    SWIFT J0001.6-7701       & Fairall 1203                & 	 0.4419          & -76.9540    & 0.0584   & 4          &  LOB     \\
    SWIFT J0005.0+7021       & 2MASX J00040192+7019185     &   1.0082          & 70.3218     & 0.0960   & 5          &  LOB     \\
    SWIFT J0010.5+1057       & Mrk 1501                    &   2.6292          & 	10.9749    & 0.0893   & 4          &  UNOB    \\
    SWIFT J0034.5-7904       & 2MASX J00341665-7905204     &   8.5698          & -79.0890    & 0.0740   & 4          &  UNOB    \\
		SWIFT J0048.8+3155$^\dagger$ & Mrk 348                 &  12.1964          &  31.9570    & 0.0150   & 5          &  MOB     \\
    SWIFT J0051.9+1724       & Mrk 1148                    &  12.9782          &  17.4329    & 0.0640   & 4          &  UNOB    \\
\hline  
	\end{tabular}
\end{table*}

\addtocounter{table}{-1}	

\begin{table*}
	\centering
	\caption{Continued.}
	\label{tab:best_fit}
	\begin{tabular}{llrrrcr} 
		\hline
		{Source Name}                & Counter part      &  RA                       &Dec             &z      &BAT Class &$N_H$ class    \\
		                             &                   &                           &                &       &          &               \\
		\hline	  
	  	  
    SWIFT J0241.6+0711       & Mrk 595                     &  40.3954          &   7.1872    & 0.0270   & 4          &  LOB     \\
    SWIFT J0244.8+6227       & [HB89] 0241+622             &  41.2404          &  62.4685    & 0.0440   & 4          &  UNOB    \\
    SWIFT J0308.5-7251       & ESO 031- G 008              &  46.8972          & -72.8340    & 0.0276   & 4          &  UNOB    \\
		SWIFT J0318.7+6828$^\dagger$ & 2MASX J03181899+6829322 &  49.5791          &  68.4921    & 0.0901   & 5          &  LOB     \\
    SWIFT J0324.8+3410$^\dagger$ & 1H 0323+342             & 51.1715           &  34.1794    & 0.0610   & 4          &  UNOB    \\
    SWIFT J0325.0-4154       & LCRS B032315.2-420449       &  51.2593          & -41.9049    & 0.0580   & 4          &  UNOB    \\
\hline  
	\end{tabular}
\end{table*}

\addtocounter{table}{-1}	

\begin{table*}
	\centering
	\caption{Continued.}
	\label{tab:best_fit}
	\begin{tabular}{llrrrcr} 
		\hline
		{Source Name}                & Counter part      &  RA                       &Dec             &z      &BAT Class &$N_H$ class    \\
		                             &                   &                           &                &       &          &               \\
		\hline	  
	  
    SWIFT J1044.1+7024       & MCG +12-10-067              & 161.0356          & 70.4054     & 0.0336   & 5          &  MOB     \\
	 	SWIFT J1044.8+3812       & MCG +06-24-008              & 161.2040          & 38.1810     & 0.0259   & 5          &  LOB     \\
	 	SWIFT J1046.8+2556       & UGC 05881                   & 161.6772          & 25.9316     & 0.0206   & 5          &  MOB     \\
		SWIFT J1049.4+2258$^\dagger$ & Mrk 417                 & 162.3789          &  22.9646    & 0.0328   & 5          &  MOB     \\
    SWIFT J1052.8+1043       & 2MASX J10523297+1036205     & 163.1374          &  10.6056    & 0.0878   & 4          &  LOB     \\
	 	SWIFT J1059.8+6507       & 2MASX J10594361+6504063     & 164.9315          & 65.0684     & 0.0836   & 5          &  LOB     \\
\hline  
	\end{tabular}
\end{table*}

\addtocounter{table}{-1}	

\begin{table*}
	\centering
	\caption{Continued.}
	\label{tab:best_fit}
	\begin{tabular}{llrrrcr} 
		\hline
		{Source Name}                & Counter part      &  RA                       &Dec             &z      &BAT Class &$N_H$ class    \\
		                             &                   &                           &                &       &          &               \\
		\hline	  
	  
    SWIFT J1457.8-4308$^\dagger$ & IC 4518A                & 224.4288          & -43.1317    & 0.0157   & 5          &  MOB     \\
		SWIFT J1504.2+1025$^\dagger$ & Mrk 841                 & 226.0050          &  10.4378    & 0.0364   & 4          &  UNOB    \\		
    SWIFT J1506.7+0353A      & Mrk 1392                    & 226.4856          &   3.7073    & 0.0361   & 4          &  UNOB    \\
	 	SWIFT J1506.7+0353B      & 2MASX J15064412+0351444     & 226.6839          &  3.8623     & 0.0373   & 5          &  UNOB    \\
    SWIFT J1513.8-8125       & 2MASX J15144217-8123377     & 228.6751          & -81.3939    & 0.0684   & 4          &  LOB     \\
	 	SWIFT J1515.0+4205       & NGC 5899                    & 228.7634          & 42.0498     & 0.0085   & 5          &  MOB     \\
\hline  
	\end{tabular}
\end{table*}

\addtocounter{table}{-1}	

\begin{table*}
	\centering
	\caption{Continued.}
	\label{tab:best_fit}
	\begin{tabular}{llrrrcr} 
		\hline
		{Source Name}                & Counter part      &  RA                       &Dec             &z      &BAT Class &$N_H$ class    \\
		                             &                   &                           &                &       &          &               \\
		\hline	  
	    
	  SWIFT J2134.9-2729       & 2MASX J21344509-2725557     & 323.6880          & -27.4321    & 0.0690   & 4          &  UNOB    \\
		SWIFT J2200.9+1032$^\dagger$ & Mrk 520                 & 330.1746          &  10.5497    & 0.0266   & 5          &  UNOB    \\
		SWIFT J2201.9-3152$^\dagger$ & NGC 7172                & 330.5079          & -31.8697    & 0.0087   & 5          &  MOB     \\
    SWIFT J2209.4-4711       & NGC 7213                    & 332.3180          & -47.1666    & 0.0058   & 4          &  UNOB    \\
    SWIFT J2211.7+1843       & II Zw 171                   & 332.9745          &  18.6972    & 0.0700   & 4          &  UNOB    \\
    SWIFT J2223.9-0207       & 3C 445                      & 335.9564          &  -2.1036    & 0.0559   & 5          &  MOB     \\
\hline  
	\end{tabular}
	\tablefoot{The name and counter part of each source, its celestial coordinates as well as its BAT class, are as given in the BAT catalogue. The redshift was retrieved by NED (\textit{https://ned.ipac.caltech.edu/}) and SIMBAD (\textit{http://simbad.u-strasbg.fr/simbad/}). The last column lists the $N_H$ based classification of each object. We have defined three groups; the unobscured ($N_H< 5 \cdot 10^{22} cm^{-2}$, UNOB), the lightly obscured ($5 \cdot 10^{22} cm^{-2} < N_H < 10^{23} cm^{-2}$, LOB) and the mildly obscured ($N_H >  10^{23} cm^{-2}$, MOB) class. Sources with moderate spectral variability have not been considered in this classification.}
\end{table*}

\begin{table*}
	\centering
	\caption{Observation log of the used sample.}
	\label{tab:obs_log}
\begin{tabular}{lrrr} 
		\hline
		{Source Name}                &NuSTAR Obs ID &Obs date      &Exposure      \\
		                             &              & (MJD)        &(ks)           \\
		\hline
    SWIFT J0001.6-7701      & 60160002002    & 57123.549    & 33.91         \\
    SWIFT J0005.0+7021      & 60061001002    & 58463.427    & 23.60         \\
    SWIFT J0010.5+1057      & 60301014002    & 58098.661    & 73.05         \\
    SWIFT J0034.5-7904      & 60160015002    & 57475.793    & 22.60         \\
		SWIFT J0048.8+3155      & 60160026002    & 57323.296    & 21.52         \\
    SWIFT J0051.9+1724      & 60160028002    & 58132.618    & 17.17         \\
\hline  
	\end{tabular}
\end{table*}

\addtocounter{table}{-1}	

\begin{table*}
	\centering
	\caption{Continued.}
	\label{tab:obs_log}
	\begin{tabular}{lrrr} 
		\hline
		{Source Name}                &NuSTAR Obs ID        &Obs date      &Exposure      \\
		                             &              & (MJD)        &(ks)           \\
		\hline	  
	  
	  SWIFT J0241.6+0711      & 60160119002    & 57771.355    & 21.30         \\
    SWIFT J0244.8+6227      & 60160125002    & 57600.282    & 22.12         \\
    SWIFT J0308.5-7251      & 60160141002    & 57866.654    & 25.05         \\
		SWIFT J0318.7+6828      & 60061342002    & 57512.802    & 24.04         \\
    SWIFT J0324.8+3410      & 60061360002    & 56731.742    & 100.95        \\
    SWIFT J0325.0-4154      & 60160152002    & 57374.278    & 21.86         \\
\hline  
	\end{tabular}
\end{table*}

\addtocounter{table}{-1}	

\begin{table*}
	\centering
	\caption{Continued.}
	\label{tab:obs_log}
	\begin{tabular}{lrrr} 
		\hline
		{Source Name}                &NuSTAR Obs ID        &Obs date      &Exposure      \\
		                             &              & (MJD)        &(ks)           \\
		\hline	  
	  
    SWIFT J0917.2-6221      & 90401630002    & 58361.030    & 111.72        \\
    SWIFT J0918.5+1618      & 60061090002    & 57019.562    & 21.52         \\
	 	SWIFT J0920.1+3712      & 60061353002    & 56821.979    & 32.78         \\
		SWIFT J0920.8-0805      & 60061091002    & 56385.532    & 12.38         \\   
		                        & 60061091004    & 56392.605    &  9.39         \\  
		                        & 60061091006    & 56400.693    & 11.59         \\
\hline  
	\end{tabular}
\end{table*}

\addtocounter{table}{-1}	

\begin{table*}
	\centering
	\caption{Continued.}
	\label{tab:obs_log}
	\begin{tabular}{lrrr} 
		\hline
		{Source Name}                &NuSTAR Obs ID        &Obs date      &Exposure      \\
		                             &              & (MJD)        &(ks)           \\
		\hline	  
	  
    SWIFT J1145.6-1819      & 60302002002    & 58103.515    & 20.76         \\ 
                            & 60302002004    & 58105.529    & 19.19         \\   
                            & 60302002006    & 58107.408    & 22.65         \\  
                            & 60302002008    & 58109.556    & 20.59         \\ 
                            & 60302002010    & 58111.570    & 22.10         \\
    SWIFT J1149.3-0414      & 60061215002    & 57786.146    & 31.88         \\
\hline  
	\end{tabular}
\end{table*}

\addtocounter{table}{-1}	

\begin{table*}
	\centering
	\caption{Continued.}
	\label{tab:obs_log}
	\begin{tabular}{lrrr} 
		\hline
		{Source Name}                     &NuSTAR Obs ID        &Obs date      &Exposure      \\
		                                  &              & (MJD)        &(ks)           \\
		\hline	  
	    
	  SWIFT J1429.2+0118      & 60061254002    & 58340.381    & 32.48         \\
	 	SWIFT J1433.9+0528      & 60061337002    & 56848.634    & 20.49         \\ 	
    SWIFT J1436.4+5846      & 60160590002    & 57228.489    & 20.53         \\
		SWIFT J1441.4+5341      & 60061255002    & 56792.194    & 18.08         \\     
		                        & 60061255004    & 56801.208    & 16.80         \\
    SWIFT J1446.7-6416      & 60061257002    & 56440.318    & 15.04         \\
\hline  
	\end{tabular}
\end{table*}

\addtocounter{table}{-1}	

\begin{table*}
	\centering
	\caption{Continued.}
	\label{tab:obs_log}
	\begin{tabular}{lrrr} 
		\hline	                           
		{Source Name}                &NuSTAR Obs ID        &Obs date      &Exposure      \\
		                             &              & (MJD)        &(ks)           \\
		\hline
  
    SWIFT J1926.9+4140      & 60061334002    & 56798.593    & 21.35         \\
    SWIFT J1930.5+3414      & 60376001002    & 58036.289    & 50.32         \\
		SWIFT J1933.9+3258      & 60160714002    & 57539.642    & 12.65         \\
		SWIFT J1937.5-0613      & 60101003002    & 57296.750    & 65.02         \\
    SWIFT J1938.1-5108      & 60160716002    & 57584.225    & 21.83         \\
		SWIFT J1942.6-1024      & 60201028002    & 57573.751    & 143.12        \\
\hline  
	\end{tabular}
	\tablefoot{The third column lists the start date of each observation.}
\end{table*}

\begin{table*}
	\centering
	\caption{Best-fit parameters for all the sources.}
	\label{tab:best_fit}
	\begin{tabular}{lcccrrr} 
		\hline
		{Source Name}                &  $N_H (10^{22}$cm$^{-2})$   &  $\Gamma$      & $R$                             &Fe K$\alpha$ EW    &$E_c$    &$\chi^2$/df    \\
		                             &                             &                &                                         &   (eV)    & (keV)       &                \\
		\hline
		
    SWIFT J0001.6-7701    & $7.7 \pm 2.0$              & $1.88 \pm 0.15$          & $0.89^{+0.66}_{-0.50}$   & $99^{+43}_{-46}$    & $>421$   & $143.9/154$   \\
    SWIFT J0005.0+7021    & $8.6^{+1.7}_{-3.3}$        & $1.81^{+0.19}_{-0.17}$   & $0.30^{+0.65}_{-0.30}$   & $104^{+50}_{-41}$   & $>125$   & $141.0/142$   \\
    SWIFT J0010.5+1057    & $<0.2$                     & $1.81 \pm 0.03$          & $0.15^{+0.10}_{-0.08}$   & $32^{+12}_{-10}$    & $>442$   & $774.2/801$   \\
    SWIFT J0034.5-7904    & $1.1^{+1.4}_{-1.1}$        & $2.20^{+0.12}_{-0.13}$   & $0.92^{+0.65}_{-0.51}$   & $120^{+34}_{-33}$   & $>234$   & $213.0/209$   \\
    SWIFT J0048.8+3155    & $10.9 \pm 0.7$             & $1.65 \pm 0.05$          & $0.27^{+0.11}_{-0.10}$   & $58 \pm 12$         & $>270$     & $838.9/926$   \\
    SWIFT J0051.9+1724    & $0.7^{+1.1}_{-0.7}$        & $1.86^{+0.06}_{-0.07}$   & $<0.10$                  & $36^{+21}_{-19}$    & $>156$   & $433.7/437$   \\
\hline  
	\end{tabular}
\end{table*}

\addtocounter{table}{-1}	

\begin{table*}
	\centering
	\caption{Continued.}
	\label{tab:best_fit}
	\begin{tabular}{lcccccr} 
		\hline
		{Source Name}                &  $N_H (10^{22}$cm$^{-2})$   &  $\Gamma$      & $R$                             &Fe K$\alpha$ EW    &$E_c$    &$\chi^2$/df    \\
		                             &                             &                &                                         &   (eV)    & (keV)   &                \\
		\hline

    SWIFT J0714.3+4541    & $0.9^{+1.1}_{-0.9}$        & $2.11^{+0.12}_{-0.10}$   & $2.39^{+0.91}_{-0.79}$   & $136^{+40}_{-45}$   & $>911$  & $164.2/162$   \\
    SWIFT J0736.9+5846    & $<1.5$                     & $1.84^{+0.15}_{-0.11}$   & $1.37^{+0.96}_{-0.58}$   & $111 \pm 58$        & $>744$  & $115.9/112$   \\
    SWIFT J0739.7-3142    & $27.9^{+2.8}_{-2.7}$       & $1.73(f)$                & $1.86^{+0.64}_{-0.50}$   & $247^{+46}_{-49}$   & $>187$  & $146.0/163$   \\
    SWIFT J0742.3+8024    & $13.6 \pm 3.9$             & $1.48^{+0.22}_{-0.23}$   & $<0.32$                  & $<34$               & $51^{+97}_{-22}$  & $113.2/115$   \\
    SWIFT J0804.2+0507    & $24.1^{+2.5}_{-2.4}$       & $1.65 \pm 0.17$          & $1.33^{+0.69}_{-0.44}$   & $121^{+28}_{-36}$   & $143^{+286}_{-60}$   & $360.6/320$   \\
    SWIFT J0804.6+1045    & $14.6^{+2.2}_{-1.2}$       & $1.83^{+0.13}_{-0.09}$   & $0.02^{+0.32}_{-0.02}$   & $61^{+38}_{-45}$    & $>113$   & $176.3/167$   \\
\hline  
	\end{tabular}
\end{table*}

\addtocounter{table}{-1}	

\begin{table*}
	\centering
	\caption{Continued.}
	\label{tab:best_fit}
	\begin{tabular}{lcccccr} 
		\hline
		{Source Name}                &  $N_H (10^{22}$cm$^{-2})$   &  $\Gamma$      & $R$                             &Fe K$\alpha$ EW    &$E_c$    &$\chi^2$/df    \\
		                             &                             &                &                                         &   (eV)    & (keV)   &                \\
		\hline
   
    SWIFT J1139.8+3157    & $6.9^{+2.5}_{-4.5}$        & $2.12^{+0.28}_{-0.40}$   & $1.71^{+2.67}_{-1.33}$   & $<155$              &  $>50$  & $34.2/39$   \\
    SWIFT J1142.2+1021    & $<0.9$                     & $1.70^{+0.15}_{-0.12}$   & $0.30^{+0.43}_{-0.30}$   & $161^{+73}_{-65}$   &  $>88$  & $77.7/75$   \\
    SWIFT J1143.7+7942    & $0.2^{+0.7}_{-0.2}$        & $1.73^{+0.06}_{-0.04}$   & $0.49^{+0.17}_{-0.15}$   & $67^{+16}_{-15}$    &  $266^{+458}_{-92}$  & $712.7/708$   \\
    SWIFT J1145.6-1819    & $0.1^{+0.5}_{-0.1}$        & $1.74^{+0.04}_{-0.02}$   & $0.26^{+0.08}_{-0.07}$   & $96^{+15}_{-14}$    &  $112^{+24}_{-16}$   & $1108.3/1083$   \\
    SWIFT J1149.3-0414    & $<0.6$                     & $1.78^{+0.09}_{-0.07}$   & $0.16^{+0.34}_{-0.16}$   & $171^{+47}_{-44}$   &  $>208$ & $126.1/158$   \\
    SWIFT J1152.6-0512    & $0.5^{+1.3}_{-0.5}$        & $1.90^{+0.10}_{-0.09}$   & $0.33^{+0.34}_{-0.27}$   & $<24$               &  $>511$ & $175.7/206$   \\
\hline  
	\end{tabular}
\end{table*}

\addtocounter{table}{-1}	

\begin{table*}
	\centering
	\caption{Continued.}
	\label{tab:best_fit}
	\begin{tabular}{lcccccr} 
		\hline
		{Source Name}                &  $N_H (10^{22}$cm$^{-2})$   &  $\Gamma$      & $R$                             &Fe K$\alpha$ EW    &$E_c$    &$\chi^2$/df    \\
		                             &                             &                &                                         &   (eV)    & (keV)   &                \\
		\hline
    
    SWIFT J1515.0+4205    & $10.9 \pm 1.0$             & $1.89^{+0.05}_{-0.07}$   & $0.89^{+0.41}_{-0.37}$   & $113^{+34}_{-33}$   &  $>705$  & $282.7/248$   \\
    SWIFT J1519.6+6538    & $21.7^{+3.8}_{-3.7}$       & $1.73(f)$                & $0.98^{+0.55}_{-0.35}$   & $<66$               &  $>246$  & $65.0/52$   \\
    SWIFT J1530.0-1300    & $2.0^{+1.8}_{-1.9}$        & $1.90^{+0.13}_{-0.12}$   & $0.57^{+0.40}_{-0.36}$   & $67^{+34}_{-36}$    &  $>492$  & $190.6/189$   \\
    SWIFT J1533.2-0836    & $5.8^{+1.1}_{-1.9}$        & $1.67^{+0.03}_{-0.13}$   & $0.34^{+0.55}_{-0.32}$   & $185^{+40}_{-53}$   &  $>291$  & $144.3/165$   \\
    SWIFT J1535.9+5751    & $0.4^{+1.1}_{-0.4}$        & $1.72^{+0.10}_{-0.06}$   & $0.60^{+0.30}_{-0.24}$   & $104^{+24}_{-25}$   &  $225^{+946}_{-105}$  & $431.7/478$   \\
    SWIFT J1548.5-1344    & $1.9 \pm 0.8$              & $2.02^{+0.08}_{-0.07}$   & $1.00^{+0.39}_{-0.31}$   & $139 \pm 21$        &  $>1393$  & $456.7/458$   \\
\hline  
	\end{tabular}
\end{table*}

\addtocounter{table}{-1}	

\begin{table*}
	\centering
	\caption{Continued.}
	\label{tab:best_fit}
	\begin{tabular}{lcccccr} 
		\hline
		{Source Name}                &  $N_H (10^{22}$cm$^{-2})$   &  $\Gamma$      & $R$                             &Fe K$\alpha$ EW    &$E_c$    &$\chi^2$/df    \\
		                             &                             &                &                                         &   (eV)    & (keV)   &                \\
		\hline
		
		SWIFT J2209.4-4711    & $1.2^{+0.4}_{-0.2}$        & $1.87 \pm 0.04$          & $<0.06$                  & $84^{+14}_{-11}$    &  $340^{+165}_{-159}$  & $832.7/883$   \\
    SWIFT J2211.7+1843    & $0.7^{+0.8}_{-0.7}$        & $1.93 \pm 0.06$          & $0.41^{+0.18}_{-0.16}$   & $25^{+14}_{-12}$    &  $155^{+150}_{-53}$  & $752.1/677$   \\
    SWIFT J2223.9-0207    & $21.1^{+1.8}_{-1.7}$       & $1.73(f)$                & $1.16^{+0.38}_{-0.32}$   & $64 \pm 35$         &  $142^{+103}_{-43}$  & $261.2/247$   \\
    SWIFT J2235.9-2602    & $1.1 \pm 0.2$              & $2.09^{+0.01}_{-0.02}$   & $1.09^{+0.12}_{-0.11}$   & $170 \pm 13$        &  $>3190$  & $1233.4/1222$   \\
    SWIFT J2236.7-1233    & $4.5^{+0.5}_{-0.8}$        & $1.87 \pm 0.03$          & $0.51 \pm 0.19$          & $90^{+19}_{-15}$    &  $>1408$  & $606.0/602$   \\  
                          & $2.5 \pm 1.3$              & $1.55 \pm 0.10$          & $<0.22$                  & $140^{+23}_{-26}$   &  $172^{+689}_{-78}$  & $473.4/446$   \\
\hline  
	\end{tabular}
	\tablefoot{The first column lists the Swift BAT name of the source. The next three columns list the best-fit results for the absorption, spectral index and reflection strength. The fifth column lists the Fe line equivalent width as estimated from the best-fit model and the sixth column lists the best-fit values of the high-energy cutoff. The last column lists the fit $\chi^2$ statistic and the degrees of freedom. The error on equivalent width was estimated with $N_H$ and $R$ being fixed to their best-fit values and thus is only an underestimation of the real error. The values of $1.73(f)$ in the third column denote the sources that were fit with the power-law index fixed. The upper limits given in the second, fourth and fifth column and the lower limits in the sixth column correspond to 1-$\sigma$ level.}
\end{table*}

\end{document}